\documentclass[12pt]{article}

\pdfoutput=1

\usepackage[inner=1.5cm,outer=4cm]{geometry}

\usepackage{cite}

\usepackage{amsmath,amssymb,latexsym}

\usepackage[T1]{fontenc}
\usepackage{times}

\usepackage{epstopdf}

\usepackage{graphics,epsfig}

\usepackage{color}

\usepackage{mathrsfs}

\usepackage{fdsymbol}

\DeclareMathAlphabet{\mathpzc}{OT1}{pzc}{m}{it}

\let\a=\alpha \let\b=\beta \let\g=\gamma \let\d=\delta \let\e=\epsilon
\let\z=\zeta  \let\th=\theta  \let\k=\kappa
\let\l=\lambda \let\m=\mu \let\n=\nu \let\x=\xi \let\p=\pi 
\let\s=\sigma \let\t=\tau  \let\f=\phi  \let\y=\psi
 
\let\w=\omega      \let\G=\Gamma  \let\Th=\Theta \let\L=\Lambda
\let\X=\Xi  \let\S=\Sigma  \let\Y=\Psi
 
\let\la=\label  
  
 \def\bd{\begin{document}} \def\ed{\end{document}}
\def\ds{\documentstyle} \let\fr=\frac \let\bl=\bigl \let\br=\bigr
\let\Br=\Bigr \let\Bl=\Bigl
\let\bm=\bibitem
\let\na=\nabla
\def\tU{{\widetilde U}}
\let\pa=\partial \let\ov=\overline
\def\ie{{\it i.e.\ }}
\newcommand{\be}{\begin{equation}}
\newcommand{\ee}{\end{equation}}
\def\ba{\begin{array}}
\def\ea{\end{array}}
\def\ft#1#2{{\textstyle{{\scriptstyle #1}\over {\scriptstyle #2}}}}
\def\fft#1#2{{#1 \over #2}}
\def\F#1#2{{ F_{#1}^{(#2)} }}
\def\cF#1#2{{ {\cal F}_{#1}^{(#2)} }}

\def\R{{\bf R}}
\def\sst#1{{\scriptscriptstyle #1}}
\def\oneone{\rlap 1\mkern4mu{\rm l}}
\def\e7{E_{7(+7)}}
\def\td{\tilde}
\def\wtd{\widetilde}
\def\im{{\rm i}}
\def\bog{Bogomol'nyi\ }
\newcommand{\ho}[1]{$\, ^{#1}$}
\newcommand{\hoch}[1]{$\, ^{#1}$}
\newcommand{\bea}{\begin{eqnarray}}
\newcommand{\eea}{\end{eqnarray}}
\newcommand{\ra}{\rightarrow}
\newcommand{\lra}{\longrightarrow}
\newcommand{\Lra}{\Leftrightarrow}
\newcommand{\ap}{\alpha^\prime}
\newcommand{\bp}{\tilde \beta^\prime}
\newcommand{\cB}{{\cal B}}
\newcommand{\cO}{{\cal O}}
\newcommand{\vecx}{\vec{x}}
\newcommand{\vecy}{\vec{y}}
\newcommand{\vecp}{\vec{p}}
\newcommand{\vecq}{\vec{q}}
\newcommand{\tr}{{\rm tr} }
\newcommand{\Tr}{{\rm Tr} }
\newcommand{\NP}{Nucl. Phys. }

\newcommand{\cL}{{\cal L}}
\newcommand{\cA}{{\cal A}}
\newcommand{\cT}{{\cal T}}
\newcommand{\cR}{{\cal R}}
\newcommand{\cD}{{\cal D}}
\newcommand{\cH}{{\cal H}}

\def\Cb{\bar{C}}

\def\sst#1{{\scriptscriptstyle #1}}
\def\0{{\sst{(0)}}}
\def\1{{\sst{(1)}}}
\def\2{{\sst{(2)}}}
\def\3{{\sst{(3)}}}
\def\4{{\sst{(4)}}}
\def\5{{\sst{(5)}}}
\def\6{{\sst{(6)}}}
\def\7{{\sst{(7)}}}
\def\8{{\sst{(8)}}}
\def\9{{\sst{(9)}}}
\def\p{{\sst{(p)}}}
\def\q{{\sst{(q)}}}
\def\ve{\varepsilon}
\def\vf{\varphi}
\def\F{\Phi}
\def\wg{\wedge}

\def\thb{\bar{\theta}}
\def\Thb{\bar{\Theta}}
\def\barp{\bar{p}}
\def\barq{\bar{q}}
\def\barc{\bar{c}}
\def\bard{\bar{d}}
\def\e{\epsilon}

\def \bi{\bibitem}
\def \la {\label}

\def \l {\lambda}
\def\foot{\footnote}
\def \tl  {{\tilde \l}}
\def \sql {{\sqrt \l}}
\def \adss {$AdS_5 \times S^5$\ }
\newcommand{\rf}[1]{(\ref{#1})}
\def \ov {\over}

\def\th{\theta}
\def\Th{\Theta}
\def\vth{\vartheta}
\def\btheta{{\bar\theta}}
\def\ttheta{{{\tilde\theta}}}
\def\bttheta{{{\bar\ttheta}}}
\def\vth{\vartheta}

\def\ra{\rightarrow}
\def\N{\nabla}
\def\F{{\cal F}}
\def\uM{\underline{M}}
\def\uA{\underline{A}}
\def\uN{\underline{N}}
\def\uP{\underline{P}}
\def\ua{\underline{a}}
\def\ub{\underline{b}}
\def\uc{\underline{c}}
\def\ud{\underline{d}}
\def\ue{\underline{e}}
\def\uf{\underline{f}}
\def\ui{\underline{i}}
\def\uj{\underline{j}}
\def\uk{\underline{k}}
\def\ul{\underline{l}}
\def\ual{\underline{\alpha}}
\def\ube{\underline{\beta}}
\def\um{\underline{m}}
\def\un{\underline{n}}
\def\up{\underline{p}}
\def\uq{\underline{q}}
\def\ur{\underline{r}}
\def\us{\underline{s}}
\def\umu{\underline{\mu}}
\def\unu{\underline{\nu}}
\def\ula{\underline{\l}}
\def\uka{\underline{\k}}
\def\usi{\underline{\s}}
\def\urh{\underline{\r}}
\def\cc{\circ}
\def\eqv{\equiv}

\def\ni{\noindent}

\def\Ep{E^{{}^{(+)}}}
\def\Em{E^{{}^{(-)}}}

\def\Mp{M^{{}^{(+)}}}
\def\Mm{M^{{}^{(-)}}}

\def \ha{{1\ov 2}}

\def\r{\rho}

\def\Y{{\rm Y}}
\def\X{{\rm X}}
\def\tY{\tilde{\rm Y}}
\def\tX{\tilde{\rm X}}
\def\dY{\dot{\rm Y}}
\def\dX{\dot{\rm X}}

\def \J {\mathcal{J}}
\def \del {\partial}

\def\dF{\dot{F}}
\def\dG{\dot{G}}
\def\df{\dot{f}}
\def \E {{\cal E}}
\def \S {{\cal S}}
\def \J {{\cal J}}

\def\ms{\mathcal{S}}
\def\mj{\mathcal{J}}
\def\soj{\fr{\ms}{\mj}}
\def \R {{\bf R}}
\def \om {\omega}
\def \bE {\bar E}
\def \x {{\cal X}}

\def \bi{\bibitem}
\def \la {\label}

\def \l {\lambda}
\def\foot{\footnote}
\def \tl  {{\tilde \l}}
\def \sql {{\sqrt \l}}
\def \adss {$AdS_5 \times S^5$\ }
\def \ov {\over}

\def \varpi {{\rm w}}

\def\thb{\bar{\theta}}
\def\Thb{\bar{\Theta}}
\def\mb{\bar{\m}}
\def\ab{\bar{\a}}
\def\zb{\bar{z}}
\def\psib{\bar{\psi}}
\def\barp{\bar{p}}
\def\barq{\bar{q}}
\def\barc{\bar{c}}
\def\bard{\bar{d}}
\def\e{\epsilon}
\def\wb{\bar{w}}
\def\lb{\bar{\l}}
\def\Jb{\bar{J}}
\def\Nb{\bar{N}}
\def\Zb{\bar{Z}}
\def\pab{\bar{\pa}}

\def\At{\tilde{A}}
\def\Bt{\tilde{B}}
\def\Ct{\tilde{C}}
\def\Dt{\tilde{D}}
\def\Et{\tilde{E}}
\def\Ft{\tilde{F}}
\def\Gt{\tilde{G}}
\def\Ht{\tilde{H}}
\def\Kt{\tilde{K}}
\def\Mt{\tilde{M}}
\def\Nt{\tilde{N}}
\def\Rt{\tilde{R}}
\def\at{\tilde{a}}
\def\bt{\tilde{b}}
\def\ct{\tilde{c}}
\def\dt{\tilde{d}}
\def\et{\tilde{e}}
\def\ft{\tilde{f}}
\def\htil{\tilde{h}}
\def\gt{\tilde{g}}
\def\nt{\tilde{n}}
\def\mut{\tilde{\mu}}
\def\nut{\tilde{\nu}}
\def\pht{\tilde{\f}}
\def\Pht{\tilde{\Phi}}
\def\vft{\tilde{\vf}}
\def \zet{\tilde{\z}}

\def\rht{\tilde{\rho}}

\def\asth{\hat{*}}
\def\phh{\hat{\phi}}

\def\bA{{\bf A}}

\def\ola{\overleftarrow}
\def\ora{\overrightarrow}
\def\alt{\tilde{\a}}

\def\eh{\hat{e}}
\def\eph{\hat{\e}}
\def\ph{\hat{p}}
\def\alh{\hat{\a}}
\def\beh{\hat{\b}}
\def\gah{\hat{\g}}
\def\Fh{\hat{F}}
\def\muh{\hat{\m}}
\def\nuh{\hat{\n}}
\def\thh{\hat{\th}}
\def\rhh{\hat{\r}}
\def\dh{\hat{d}}
\def\ih{\hat{i}}
\def\jh{\hat{j}}
\def\hh{\hat{h}}
\def\nh{\hat{n}}
\def\gh{\hat{g}}
\def\kh{\hat{k}}
\def\deh{\hat{\d}}
\def\wh{\hat{w}}
\def\lah{\hat{\l}}
\def\Ah{\hat{A}}
\def\Kh{\hat{K}}
\def\Nh{\hat{N}}
\def\Rh{\hat{R}}
\def\Ch{\hat{C}}
\def\Omh{\hat{\Omega}}

\def\xh{\hat{x}}

\def\ps{\rlap{\, /}\;\,p }
\def\ks{\rlap{\, /}\;\,k }

\def\gym{g_{YM}}

\def\adot{\dot{a}}
\def\bdot{\dot{b}}
\def\bpa{\bar{\pa}}

\def\pr{\prime}
\def\ssk{\medskip}
\def\clb{\color{blue}}
\def\clr{\color{red}}
\def\clg{\color{green}}

\def\bfA{{\bf A}}
\def\bfB{{\bf B}}
\def\bfK{{\bf K}}
\def\bfU{{\bf U}}
\def\bfX{{\bf X}}
\def\bfY{{\bf Y}}
\def\bfZ{{\bf Z}}
\def\bfg{{\bf g}}
\def\bfn{{\bf n}}

\def \vk{\vec{k}}
\def \vx{\vec{x}}

\def\bsk{\bigskip}
\def\ssk{\medskip}

\def\Ec{{\cal E}}

\begin{document}

\overfullrule=0pt
\parskip=2pt
\parindent=12pt
\headheight=0in \headsep=0in \topmargin=0in
\oddsidemargin=0.45in

\vspace{ -3cm}
\thispagestyle{empty}

\vspace{0.1cm}

\setcounter{equation}{0}
\setcounter{footnote}{0}
\setcounter{section}{0}

\title{
{\Large{\bf 
Hallmarks of tunneling dynamics with broken reflective symmetry}}
}
{\vspace{-1.5cm}
\author{V.P. Berezovoj$\,^{\spadesuit}$\footnote{berezovoj@kipt.kharkov.ua}\,, M.I. Konchatnij$\,^{\spadesuit}$\footnote{konchatnij@kipt.kharkov.ua}
\,and  \,A.J. Nurmagambetov$\,^{\spadesuit,\vardiamondsuit,\varheartsuit}$\footnote{ajn@kipt.kharkov.ua}
\\ \\
$\,^{\spadesuit}${ \it {\normalsize Akhiezer Institute for Theoretical Physics of NSC KIPT}}\\
{ \it {\normalsize 1 Akademicheskaya St., Kharkov 61108 UA} }
\\
$\,^{\vardiamondsuit}${ \it {\normalsize Department of Physics \& Technology, Karazin Kharkov National University,}}\\
{ \it {\normalsize 4 Svobody Sq., Kharkov 61022 UA} }
\\
$\,^{\varheartsuit}${ \it {\normalsize Usikov Institute of Radiophysics and Electronics}}\\
{ \it {\normalsize 12 Ak. Proskury St., Kharkov 61085 UA} }
}


\date{}

\maketitle

\vspace{-1cm}
\begin{abstract}
We study features of tunneling dynamics in an exactly-solvable model of N=4 supersymmetric quantum mechanics with a multi-well potential and with broken reflective symmetry. Quantum systems with a phenomenological potential of this type demonstrate the phenomenon of partial localization of under-barrier states, possibly resulting in the appearance of the so-called ``resonant'' tunneling, or the phenomenon of  coherent tunneling destruction, referring to the complete localization. Taking the partial localization and the coherent tunneling destruction as basic examples, we indicate main advantages of using isospectral exactly-solvable Hamiltonians in studies quantum mechanical systems with two- and three-well potentials. They, in particular, are: having enough freedom of changing the potential shape in a wide range, that allows one to choose an exactly-solvable model close to characteristics of the phenomenological one; ability of changing the number of local minima and symmetry characteristics  of the potential (symmetric or deformed) without changing the main part of the spectrum; engaging a smart basis of states, that dramatically decreases the dimensionality of matrices used in the diagonalization procedure of the corresponding spectral problem.

\end{abstract}


\section{Introduction}

Tunneling is perhaps the most fascinating, enigmatic and controversial feature of the quantum world. It came almost a century since this phenomenon was discovered by Hund, Gamow, Oppenheimer, Condon and Gurney  \cite{Hund:1927-1,Hund:1927-3,Gamow:1928,Oppenheimer:1928,Gurney:1928}, but we do not have the comprehensive picture of this process yet and the ultimate conclusions on mechanisms behind. For instance, one of the controversies, related to the tunneling, is a super-luminal passing speed of a quantum object through a potential barrier \cite{Davies:2005,Winful:2006,Aichmann,Ramos:2019,Spierings:2020} (and refs. therein). Another intriguing concern is the possibility to tunnel a macroscopic structure. This issue was risen up in late 80th in different contexts (see, e.g., refs. \cite{Leggett:1987,Chudnovsky:1988,Chudnovsky:Book,Miyazaki:Book,Ankerhold:Book} for reviews), including the macroscopic tunneling in SQUIDs, magnetic domains relaxation to the ground state, atomic and molecular processes in chemical physics, biology and so on. The first experimental evidence of these theoretical ideas comes from pioneering works on the quantum tunneling of magnetization \cite{Awschalom:1992,Sangregorio:1997}; the realization of Bose-Einstein condensates of cold atoms has simplified the observation of the macroscopic quantum tunneling in all forms \cite{Shukla:2020} (and refs. therein). Another feature of the tunneling effect worth to mention is that experiments in this branch put even foundations of Quantum Mechanics \cite{Laloe:Book} on the test, since some of the tunneling properties in modern artificial materials like graphene \cite{Pandey:2019} are naturally described within alternative -- the Bohmian -- interpretation of Quantum Mechanics. Moreover, in the context of the main subject of the paper, it is important to note the role of supersymmetry in explaining the prominent features of carbon nanostructures such as the perfect Klein tunneling and the no-backscattering effect \cite{Jakubsky:2010bg}.

\ssk
Progress in experimental studies of the condensate of atoms, mentioned insofar, resulted in the foundation and development of a new discipline~-- quantum engineering~-- which currently is considered as the base of developing Q-bits, quantum transistors, new quantum-type flash-memory modules and other parts of outcoming quantum computers. We have to emphasize that the quantum engineering means not just the creation of new type devices, but, what is more important, of quantum-mechanical objects with predefined properties, together with the possibility to control these properties with external EM fields (laser pulses).  The tunneling dynamics will certainly have a profound effect on the design of quantum computer components, and ultimately, on the computing power of the final gadget. It is also interesting to emphasize the potential importance of supersymmetry for quantum computation and quantum information, pointed up in recent work \cite{Crichigno:2020vue}.

\ssk
Evidently, any effects of the tunneling dynamics disclose the specific of the potential of a quantum-mechanical model in the end. For a symmetric double-well potential it is worth emphasize the theoretically predicted phenomena of the coherent tunneling destruction (CTD) \cite{Grossmann:1991,Grossmann:1992,Longhi:2005}, and the tunneling amplification, initiated by dynamical chaos, ref. \cite{Lin:1992}, which were experimentally discovered and reported in refs. \cite{Kierig:2008,DellaValle:2007}.  The coherent tunneling destruction is the subject of numerous theoretical and experimental investigations (see, e.g., refs. \cite{Grossmann:1991ZP,Dittrich:1998Book,Llorente:1992,Kayanuma:1994,Wang:1994,Luo:2007,Lu:2010,Lu:2011,Kar:2014}) and is one of the aspects in the focus of the present work.

\ssk
Qualitatively, good enough conclusions on the tunnel dynamics can be reached with employing phenomenological models of \cite{Grossmann:1991ZP,Dittrich:1998Book,Llorente:1992,Kayanuma:1994,Wang:1994,Luo:2007,Lu:2010,Lu:2011,Ashhab:2007} (with a Higgs-type potential and its multi-well modifications with the number of wells higher than two). These results are obtained upon the assumption of a sufficiently high barrier in between wells, or, equivalently, with taking a few lowest states of the corresponding spectra. That is to say, physics of these processes is mainly determined by the lowest (under-barrier) states of Hamiltonians (by the tunnel doublet); nevertheless, the role of more higher states may be decisive. For example, a control of the tunnel dynamics with time-dependent fields is achieved in quantum mechanical models with multi-well potentials. In most cases, a multi-well potential is assigned by an analytical modification of the Higgs-type potential, that requires, in what follows, cumbersome numerical computations upon getting the (quasi)energies and the Floquet eigenstates by the diagonalization of the complete Hamiltonian. The choice of the appropriate basis of states becomes crucial in the simplification and optimization of the diagonalization procedure, since the right choice of the basis essentially reduces the dimensionality of the used in numerics matrices. From this point of view, the series expansion of states of phenomenological Hamiltonians over the Harmonic Oscillator basis (see, e.g., refs. \cite{Grossmann:1991,Grossmann:1992,Longhi:2005,Luo:2007}) is not the finest choice: it requires dealing with hundreds of basic states, that determines the dimensionality of the matrix to diagonalize.

\ssk
Another aspect in the focus of the present paper is considering tunneling processes in quantum systems with broken reflective symmetry (see refs. \cite{Nieto:1985,Bolotin:1993} for early papers in this field). The importance of studies in the tunneling dynamics of models with asymmetric multi-well potentials can be illustrated by the analysis of the control processes in a radio-frequency SQUID phase qubit \cite{Rouse:1995, Silvestrini:1996}, as well as in widely-discussed prospects of creation of the THz radiation sources \cite{Kibis:2009,Kristinsson:2014}. As we will see, shortly, the case of asymmetric potentials is more technically complicated, however encompasses new phenomena.

\ssk
Specifically, the examination of states, located in different wells of an asymmetric two-well potential \cite{Nieto:1985}, revealed the following non-trivial effect: the Gaussian wave packet in a metastable state stays at its initial localization for an infinitely long time (for an arbitrary asymmetry degree of the potential). Moreover, there is a claim, that, in a system with a symmetric two-well potential, adding a specific but small perturbation \cite{Graffi:1984}, which breaks the reflective symmetry, results in localization of under-barrier states. In ref. \cite{Landsman:2013}, on the example of a system with the two-well phenomenological potential, this claim was verified by the direct integration of the Schr\"odinger equation. Therefore, the phenomenon of localization of states in an asymmetric potential becomes an important tool of controlling the tunneling processes, in which the specific perturbation over the initially symmetric potential acts as a control parameter.

\ssk
One more branch of studies of the tunneling dynamics in systems with an arbitrary asymmetry degree of multi-well potentials is presented in refs. \cite{Dekker:1987,Dekker:1987pra,Mugnai:1988,Song:2008,Song:2011,Song:2015,Rastelli:2012,Halataei:2007,Hasegawa:2013}. First works on this subject \cite{Dekker:1987,Dekker:1987pra,Mugnai:1988} were focused on a small deformation of the potential; the value of the tunnel doublet splitting has been computed in this approximation. Further developments of refs. \cite{Song:2008,Song:2011,Song:2015,Rastelli:2012,Halataei:2007} made possible to compute the tunnel doublet levels splitting for an arbitrary deformation of the potential, as well as to find the under-barrier wave functions within the quasi-classical approximation. The Gaussian packets evolution in the model with an asymmetric two-well potential has been considered in ref. \cite{Hasegawa:2013}, where, among others, arising the resonant tunneling upon the specific values of the deformation parameters has been examined in details.

\ssk
Typically, the analysis of quantum systems with the phenomenological-like potentials is carried out either by numerical computations of the spectra and wave functions, or by use of an involved WKB computational scheme to determine the under-barrier wave functions, together with the value of the tunnel doublet splitting. The usage of multi-well potentials of a polynomial type, the asymptotic behavior of which is $x^{2n}$ (where $n$ is the number of the potential minima), faces with essential complications even for $n=2$. With growing the $n$ number, the computational schemes become more and more hard to realize. It is worth mentioning, that the aforementioned difficulties with numeric computations are related to the required high precision of the results, especially in the low part of the spectrum. Indeed, small errors in the computed energies and wave functions of the tunnel doublet may result in large deviations upon the tunneling dynamics, making possible to observe false effects. Not least at all, roots of the difficulties, one meets with, lie in the choice of the basis in the wave function series expansion upon performing the numerics. And here we turn back to what we have pointed out not so far: the choice of the Harmonic Oscillator basic states is not the optimal one, and results in growing the rank of matrices under the use.

\ssk
One of the ways to resolve difficulties, inherent to potentials of a phenomenological type, is to operate with exactly-solvable models with multi-well potentials. Examples of these models have been studied and developed in refs. \cite{Berezovoj:2010,Berezovoj:2012,Berezovoj:2013}. The advantages of the exactly-solvable models of  \cite{Berezovoj:2010,Berezovoj:2012,Berezovoj:2013} consist in: 1) having enough freedom of changing the potential shape in a wide range, that allows one to choose an exactly-solvable model close to characteristics of the phenomenological one; 2) the ability of changing the number of local minima and symmetry characteristics  of the potential (symmetric or deformed) without changing the main part of the spectrum; 3) engaging a smart basis of states, that dramatically decreases the dimensionality of matrices, used in the diagonalization procedure of the corresponding spectral problem. We will illustrate all of these benefits by different examples in below.

\ssk
The rest of the paper has the following organization.
In Section 2, we briefly discuss the machinery of construction of exactly-solvable isospectral multi-well Hamiltonians in a general case; then, we survey the results of employing this technique to the Harmonic Oscillator (HO) Hamiltonian. In Section 3 and in the following sections, we focus on the obtained Hamiltonian in different situations. Specifically, in Section 3, we describe the tunneling dynamics of states in symmetric and asymmetric two- and three-well potentials, which were obtained from the HO potential for special values of the deformation parameters. As a result, we observe a Josephson's type oscillation of the Gaussian packet in the two-well potential of a symmetric shape, and detect a partial localization of the wave packet in the case of  asymmetry of the potential. Here, we also outline the common properties of the tunneling dynamics of the Gaussian wave packet for three-well symmetric and asymmetric potentials. In Section 4, we study the response of a system with two- and three-well potentials on an external disturbance of their shapes. We compute the corresponding wave functions upon the small disturbance, far from the local minimum of the undisturbed potential, and show that the most contribution into rearranging the spectrum and the wave functions comes from seven first basic states of the exactly-solvable Hamiltonian. Checking the absence of visible changing in the spectrum and in the coefficients of series expansion of wave functions with increasing the number of basic states on ten, we verified, in this way, the efficiency of our approach. In Section 5, we examine the behavior of a quantum system with a multi-well potential upon the action of an external time-periodic driving force. We establish the criteria of the coherent tunneling destruction, and show that ten basic wave functions of the corresponding time-independent exactly-solvable Hamiltonian is good enough to describe the tunneling dynamics of the under-barrier states. Our conclusions are collected in the last section.

\ssk
{\it Notation and conventions.} We work in $\hbar=m=1$ units, recovering, if necessary, their physical values. Computations were performed for the harmonic oscillator frequency  $\w_0=1$; hence, the natural and the dimensionless coordinates $x$ and $\xi=\sqrt{\w_0} \,x$ are indistinguishable with this choice. $\w$ is reserved for the frequency of an external periodic driving force.

\section{Isospectral Hamiltonians with two- and three-well potentials: A brief survey on previous works }

There are several solution-generation methods in mathematical physics, leading to new analytic solutions of exactly-solvable differential equations upon the controlled deformation of their structure. The mentioned deformation does not intact the global structure of the initial equation (viz., its type -- elliptic or hyperbolic), but can significantly change its local form. Hence, the so specifically obtained equation will describe new dynamics, properties of which may crucially be distinct of that of the original system.

\ssk
In the context of supersymmetric quantum mechanics (SQM) one of the frequently employed solution-generation routines is the method by Crum and Krein \cite{Crum:1955,Krein:1957}. The Crum-Krein method consists of the well-defined procedure of deformation of an exactly-solvable seed Hamiltonian $H_0$ --  with the known spectrum and wave functions -- potential of which has the only one well.\footnote{Here and hereafter, $H_0$ is not just the free Hamiltonian $p^2/2$ (in $\hbar=m=1$ units); it also includes a potential $V_0$.} Specifically, this procedure is realized as the add-on, in step by step manner, new levels of the energy spectrum below the ground state of $H_0$, that results in designing a new Hamiltonian $\tilde{H}$  with a multi-well potential. The number of wells in the potential of $\tilde{H}$ is not just determined by the number of sub-ground levels; it also depends on the particular choice of parameters of the deformation (see more on that further on). The Crum-Krein method does not appeal to any particular properties of the Hamiltonians (except of their solvability), so SUSY is not a necessary prerequisite of the construction. 
However, for the supersymmetric quantum mechanical Hamiltonians, the employment of the Crum-Krein routine becomes equivalent to constructing Hamiltonians with pre-defined spectral properties within the polynomial (reducible) SQM \cite{Andrianov:1993,Andrianov:1995,Andrianov:2004}, the main attribute of which is the sequential application of the SUSY transformations to the initial Hamiltonian.

\ssk
Note, that the discussed procedure of generating a new Hamiltonian, considered in more detail in below, does not restricted to the standard exactly-solvable models of quantum mechanics and can be extended to more involved cases, when the considered system obeys a specific symmetry (like, for instance, the hidden superconformal symmetry \cite{Inzunza:2017cwh}, or the Klein four-group in superconformal mechanics \cite{Inzunza:2019xml}) and/or has a specific structure of the spectrum (as, for example, having a series of  ``valence bands'' with the own fine structure and gaps in between \cite{Carinena:2017bqs,Carinena:2017zfy}, including the periodic finite gap \cite{Correa:2008hc} or zero-gap \cite{MateosGuilarte:2017fsv}). Another interesting generalization of Supersymmetric Quantum Mechanics, realizing a correspondence of SQM to the inverse scattering method \cite{Berezovoi:1991xc}, is related to the description of multi-soliton physics \cite{Arancibia:2012zs} as well as soliton defects in a crystalline background \cite{Arancibia:2015saa} within (exotic) non-linear sypersymmetry. However, the tasks aimed in the present paper allow us to stay within the standard Crum-Krein approach, to the brief consideration of which in the cases of interest we are now moving on.

\ssk
Previously, in ref.
\cite{Berezovoj:2013}, we gained general expressions, describing a family of N=4 multi-well SQM Hamiltonians, originated from an exactly-solvable model with one-well potential, 
by adding two extra sub-levels of energies $E_{-2}<E_{-1}<E_0$ below the ground state $E_0$ of the initial Hamiltonian $H_0$. In accordance with the Crum-Krein recipe, the initial Hamiltonian gets transformed into
\be
\tilde{H}=H_0-\fr{d^2}{dx^2}\ln \fr12 \e^{JI} \mathrm{W}\{\y_{-I},\y_{-J}\},
\la{Htil}
\ee
where the indices $I,J=1,2$ count the extra sub-levels of the energy spectrum (with energies $E_{-I} \equiv (E_{-1},E_{-2})$); $\e^{IJ}$ is the 2D Levi-Civita symbol, defined as $\e^{12}=1$. Furthermore, $\y_{-I}$ are the wave functions associated with the energies $E_{-I}$ and deformation parameters $\Lambda_{-I} \in ]0,\infty[$. They are constructed out 
of linearly-independent non-normalizable solutions -- $\vf^{(A)}_{-I}(x,E_{-I})$, $A=1,2$ -- to the Schr\"odinger equations with $H_0$ at the energies $E_{-I}$ (see ref. \cite{Berezovoj:2013} for details): 
\be
\y_{-1}(x,E_{-1},\Lambda_{-1})=\vf^{(1)}_{-1}+\Lambda_{-1}\, \vf^{(2)}_{-1},
\la{psi-1def}
\ee
\be
\y_{-2}(x,E_{-2},\Lambda_{-2})=\vf^{(1)}_{-2}-\Lambda_{-2}\, \vf^{(2)}_{-2}.
\la{psi-2def}
\ee
Finally, $\mathrm{W}\{\y,\y'\}$ stands for the Wronskian of two independent solutions. The spectral problem for Hamiltonian \rf{Htil} is solved with the following set of wave functions \cite{Berezovoj:2010,Berezovoj:2012,Berezovoj:2013}:
\be
\tilde{\Psi}_{0}(x,E_{-2},\Lambda_{-2};E_{-1},\Lambda_{-1})=
\Big[\tilde{N}_{\Lambda_{-2}}\Big]^{-1} \fr{ \y_{-1}(x,E_{-1},\Lambda_{-1})}{\mathrm{W}\{\y_{-2},\y_{-1}\}},
\la{Psitil0}
\ee
\be
\tilde{\Psi}_{1}(x,E_{-1},\Lambda_{-1};E_{-2},\Lambda_{-2})=
\Big[\tilde{N}_{\Lambda_{-1}}\Big]^{-1} \fr{ \y_{-2}(x,E_{-2},\Lambda_{-2})}{\mathrm{W}\{\y_{-2},\y_{-1}\}},
\la{Psitil1}
\ee
\[
\tilde{\Psi}_i (x,E_i)=\sqrt{\fr{E_i-E_{-1}}{E_i-E_{-2}}}\,\, \times
\]
\be
\times \left( \y_i(x,E_i)+\fr{E_{-1}-E_{-2}}{E_i-E_{-2}} \fr{\mathrm{W}\{\y_i(x,E_i),\y_{-1}(x,E_{-1},\Lambda_{-1})\}}
{\mathrm{W}\{\y_{-2},\y_{-1}\}} \y_{-2}(x,E_{-2},\Lambda_{-2})
\right),
\la{Psitili}
\ee
where $\y_i(x,E_i)$, $i=0,1,\dots$ are the wave functions of the primary Hamiltonian $H_0$.

\ssk
Eqs. \rf{Htil}-\rf{Psitili} determine the general computational scheme for any exactly-solvable seed Hamiltonian. In what follows we will specify $H_0$ to be the Harmonic Oscillator (HO) Hamiltonian
\be
H_0=\fr{p^2}{2}+\fr{\w^2_0}2\,x^2.
\la{H0HO}
\ee
Then, the non-normalizable solutions at $E_{-I}<E_0$ are represented by the parabolic cylinder functions $D_\m(x)$ (see ref. \cite{Berezovoj:2013}), 
\be
\vf^{(1)}_{-1}(\xi,\bar{E}_{-1})=D_\n (\sqrt{2} \xi),\qquad
\vf^{(2)}_{-1}(\xi,\bar{E}_{-1})=D_\n (-\sqrt{2} \xi),
\la{vf1HO}
\ee
\be
\vf^{(1)}_{-2}(\xi,\bar{E}_{-2})=D_\m (\sqrt{2} \xi),\qquad
\vf^{(2)}_{-2}(\xi,\bar{E}_{-2})=D_\m (-\sqrt{2} \xi).
\la{vf2HO}
\ee
As customary in SQM computations, we have moved to the dimensionless ``coordinate'' $\xi=\sqrt{\w_0}\,x$ and the dimensionless ``energy'' $\bar{E}_{-I}=E_{-I}/\w_0$. The orders of the parabolic cylinder  functions are determined by $\n=-1/2+\bar{E}_{-1}$, $\m=-1/2+\bar{E}_{-2}$.

\ssk
The Wronskian of the linear-independent solutions becomes
\be
\mathrm{W}\,\{\vf^{(1)}_{-I},\vf^{(2)}_{-J}\}=
\fr{2\sqrt{\pi \w_0}}{\G\left(\fr{1}{2}-E_{-I}\right)}\,\d_{IJ}.
\la{Wvf12}
\ee
Consequently, the normalization constants of the newly added sub-level wave-functions \rf{Psitil0}, \rf{Psitil1} are \cite{Berezovoj:2013}
\be
\left[\tilde{N}_{\Lambda_{-2}}\right]^{-2}=4 \Lambda_{-2}\, \fr{\sqrt{\pi \w_0}}{\G(-\m)}\, (\n-\m),\quad \Lambda_{-2}>0,
\la{tilNL-2}
\ee
\be
\left[\tilde{N}_{\Lambda_{-1}}\right]^{-2}=4 \Lambda_{-1}\, \fr{\sqrt{\pi \w_0}}{\G(-\n)}\, (\n-\m),\quad \Lambda_{-1}>0.
\la{tilNL-1}
\ee
Eqs. \rf{vf1HO}-\rf{tilNL-1} empower recovering the exact expressions for the 
new Hamiltonian \rf{Htil} and the corresponding wave functions \rf{Psitil0}-\rf{Psitili} in the case of the HO seed Hamiltonian \rf{H0HO}.

\ssk
Note that multi-well potential Hamiltonians, constructed this way, inherit the property of the HO Hamiltonian -- $H(p,x)=\w_0 H(p_\xi,\xi)$ -- upon turning to the dimensionless operators of the ``coordinate'' $\xi$ and its conjugated ``momentum'' $p_\xi$. Moreover, the specific deformation procedure does not just lead to a single new Hamiltonian, but to a family of (isospectral) Hamiltonians, originated from varying the potential shape by changing the energies $\bar{E}_{-1},\bar{E}_{-2}$ and the positive deformation parameters $\Lambda_{-1},\Lambda_{-2}$. Turning to the standard $(x,p_x)$ variables back, another free parameter in hands is the frequency $\w_0$. For instance, varying $\w_0$ results in changing the positions of local minima.

\ssk
Pinning up one of the deformation parameters does not affect the generality of results, but essentially simplifies the computational scheme. We will assume hereafter (where it will not be specified apart) the unit value of the deformation parameter $\L_{-2}$, keeping variable the other parameter $\L\equiv \L_{-1}$. In addition, in all the computations below, we will assume $\w_0=1$. According to the scale property of the HO Hamiltonian, this choice corresponds to $H(p,x)=H(p_\xi,\xi)$. Therefore, we will not differentiate, in what follows, the natural and the dimensionless coordinates.

\section{Tunneling dynamics of quantum states: symmetric vs asymmetric multi-well potentials}

As we have previously mentioned, the described by \rf{Htil}-\rf{Psitili} family of Hamiltonians possesses a remarkable property: in dependence on the choice of free parameters the deformed potentials get either two or three wells. Specifically, once the tunnel doublet $\triangle=\bar{E}_{-1}-\bar{E}_{-2}$ is located far enough from the ground state of the seed Hamiltonian $\bar{E}_{0}=1/2 $, the Hamiltonian potential is shaped by a two-well curve. It turns out that the set of the parameters, resulting in two-well potentials, provides a wide ``plateau'' between the local minima (see Fig.\ref{fig:image1}). This observation makes possible to employ the two-level approximation upon studying the tunneling dynamics.

\begin{figure}[h!]
\begin{minipage}[h]{0.48\linewidth}
\center{\includegraphics[width=0.8\linewidth]{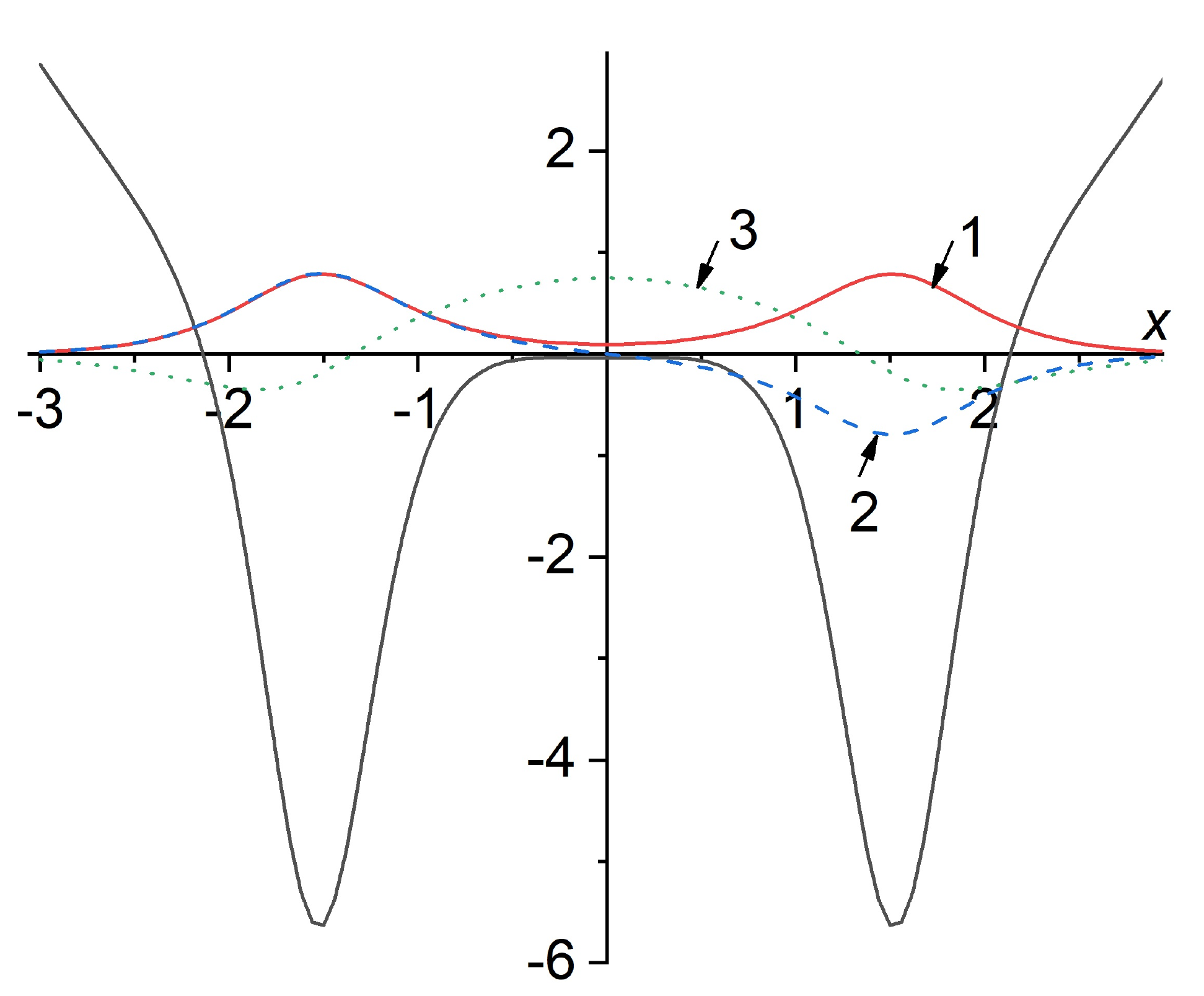} \\ a)}
\end{minipage}
\hfill
\hspace{0.99cm}
\begin{minipage}[h]{0.48\linewidth}
\center{\includegraphics[width=.8\linewidth]{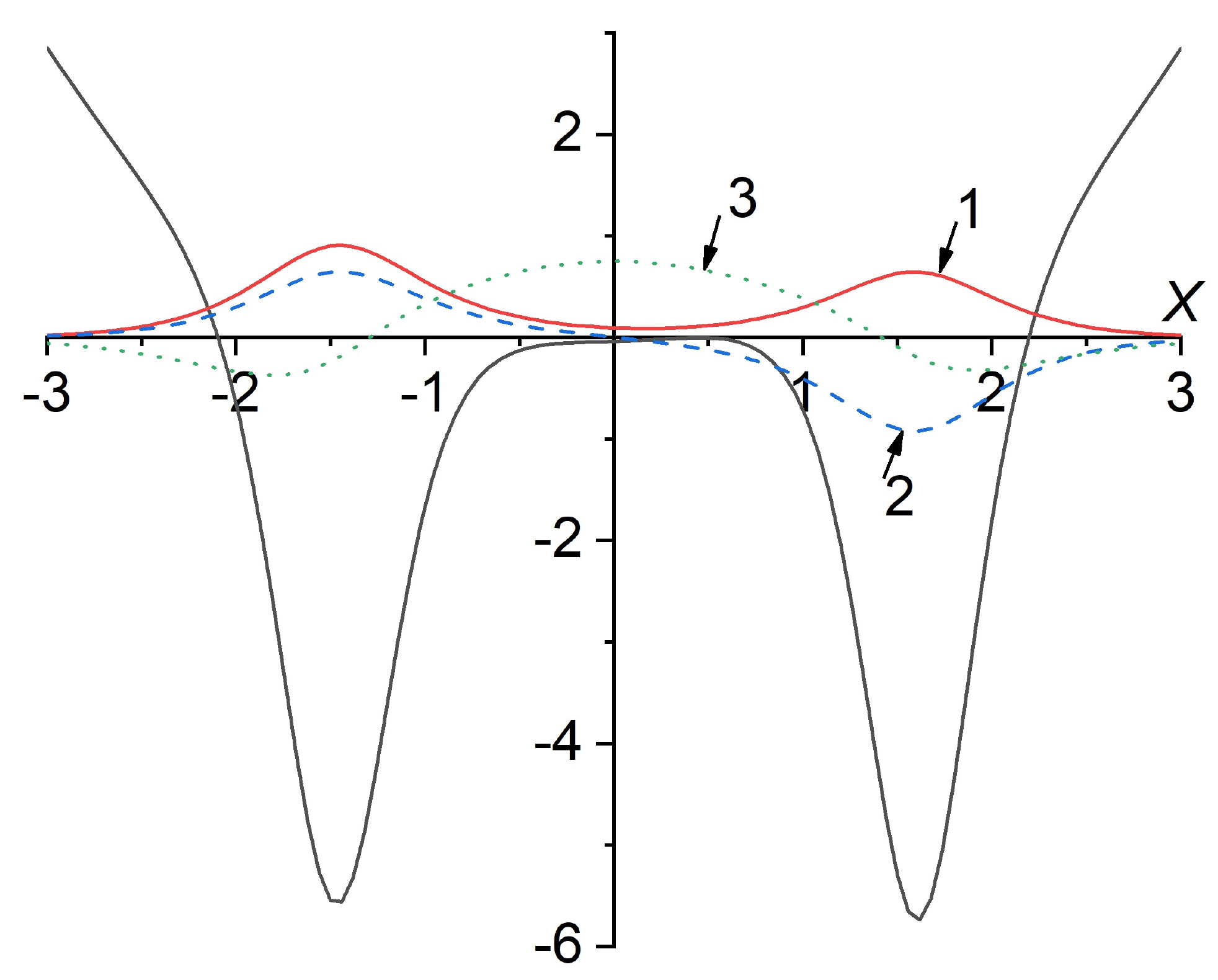} \\ b)}
\end{minipage}
\hfill
\begin{minipage}[h!]{0.99\linewidth}
\center{\includegraphics[width=0.4\linewidth]{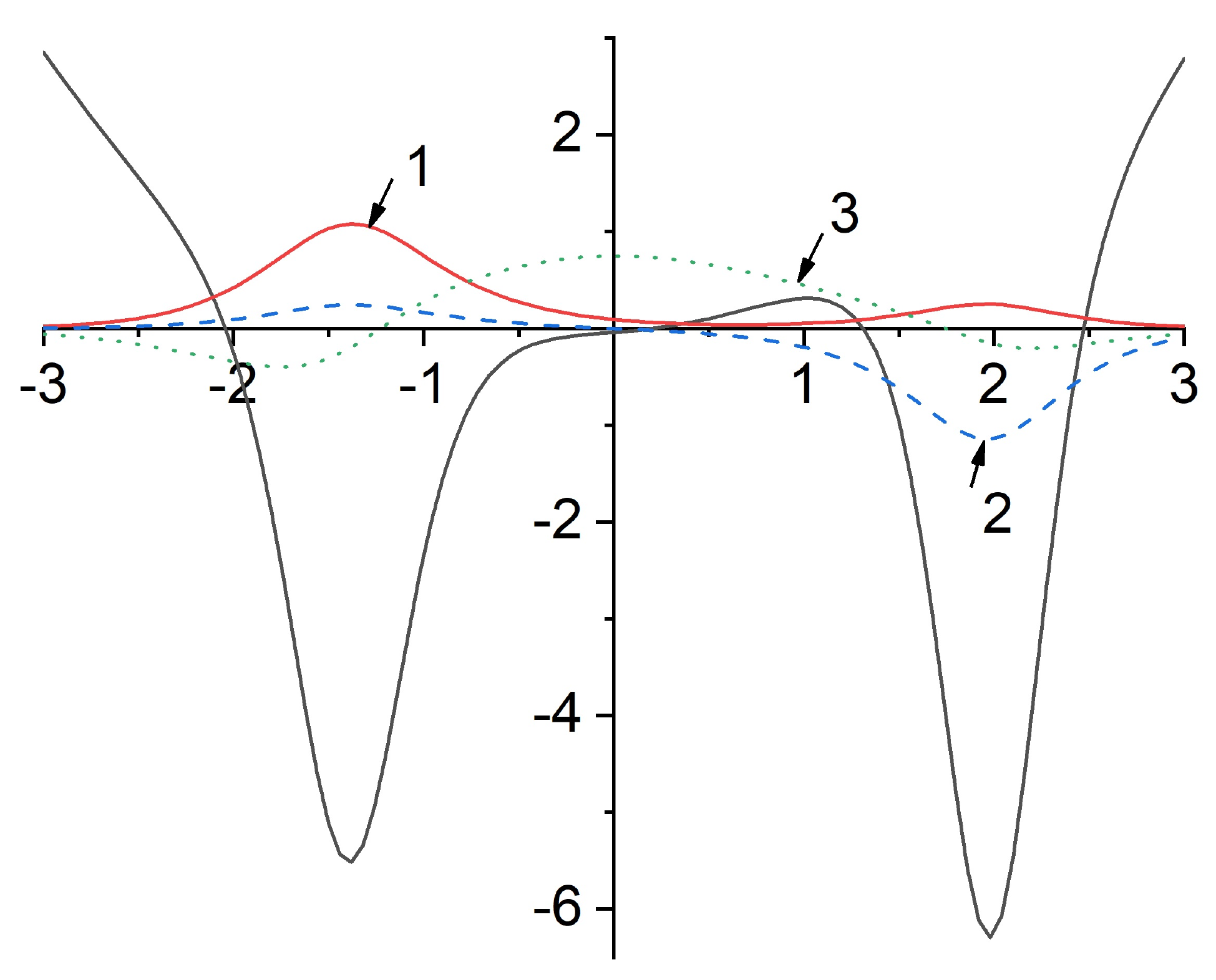} \\ c)}
\end{minipage}
\caption{Examples of two-well potentials $\tilde{U}(\xi,\bar{E}_{-2},\bar{E}_{-1},\L)$ with fixed values $\n\equiv -1/2+\bar{E}_{-1}=-3.0$, $\m\equiv -1/2+\bar{E}_{-2}=-3.02$: a) the symmetric potential ($\L=1$); b) the asymmetric potential with $\L=0.5$; c) the asymmetric potential with $\L=0.05$.  Solid lines $\bf{1}$ correspond to the ground state wave functions; dashed lines $\bf{2}$ display the first excited level wave functions; dotted lines $\bf{3}$ demonstrate the behavior of the second excited level wave functions. }
\label{fig:image1}
\end{figure}


\ssk

In the case of a symmetric potential (Fig.\ref{fig:image1}a) its shape, especially in the central part of the potential, resembles a quartic, double-well potential (the Higgs-like potentials), frequently used in the SQM literature. Tuning the value of the tunnel doublet and its location on the energy scale, one may match, with a good enough precision, the central part of the Higgs-like potential to the considered  here exactly-solvable potentials. Since the value of the tunnel doublet and wave functions for exactly-solvable Hamiltonians are explicitly known (cf. eqs. \rf{Psitil0}-\rf{tilNL-1}), they may be used as a basis in studying properties of lowest  states in models with the phenomenological interaction. Recall, the lowest states of a spectrum are responsible for the tunneling dynamics and its features.\footnote{Note that the asymptotic behavior of the phenomenological and exactly-solvable potentials differs. The asymptote of Hamiltonian \rf{Htil} is the same as the seed Hamiltonian $H_0$, while the asymptote of the Higgs-like potential is $x^{2n}$, where $n$ is the number of local minima (wells of the potential). As a result, the tunnel doublet and the wave functions under the barrier of the phenomenological type potentials are forced to change (taking into account of more and more highest states of the spectrum), that leads to essential computational difficulties.}

\ssk
Changing the value of $\L\equiv \Lambda_{-1}$ (remind, we have fixed $\Lambda_{-2}=1$), we get various potentials of a family of isospectral Hamiltonians. Decreasing the value of $\Lambda\in \,]0,\infty[$, we alter the depth of the right well. The location of its minimum shifts to higher values; the central part of the potential does not undergo essential modifications, see Fig.\ref{fig:image1}. A deformation of the potential causes the redistribution of the fraction of a wave function in each of the wells, which is the main effect. For example, the fraction of the ground state wave function increases in the left well, while the fraction of the first excited level wave function decreases therein, but increases in the right well. This effect is mostly noticeable at $\Lambda \ra 0$ (cf. Fig.\ref{fig:image1}); however, the complete localization of states does not occur.\footnote{Indeed, changing the deformation parameter does not change the spectrum of the form-invariant Hamiltonians (see ref. \cite{Berezovoj:2010} in this respect). Therefore, the value of the tunnel doublet, which characterizes the tunneling process, does not change as well. To sum up, we can only get a partial localization of the under-barrier states.}

\ssk
The partial localization of the under-barrier states gets a strong impact on the tunneling dynamics. To make this point clear, note first the viable advantage of the considered quantum Hamiltonians for studying dynamical processes: since the wave functions of the system are explicitly known (cf. eqs. \rf{Psitil0}-\rf{Psitili}), the exact propagator can also be computed analytically \cite{Berezovoj:2013}. By use of the exact propagator for the considered here HO, which acts on an originally located in the right well Gaussian wave packet,
\be
\Phi(\xi,0)=\left(\fr{e^{2R}}{\pi} \right)^{1/4} \exp\Big[{-\fr{(\xi-\xi_0)^2 e^{2R}}{2}}\Big] ,
\la{wavepack}
\ee
and choosing different sets of parameters, indicated in Fig.\ref{fig:image1}, one may observe the redistribution of the wave function fractions in two wells. Before turning to the discussion of the tunneling dynamics, caused by such redistribution, we would like to emphasize that knowing the exact propagator allows us to study (albeit technically difficult) the tunneling dynamics features of the wave packet of an arbitrary shape with taking into account all states of Hamiltonian \rf{Htil}.

\ssk
The qualitative interpretation of the wave packets localization is essentially simplified in the case when the under-barrier part of the spectrum, with $\bar{E}_n<U_{\mathrm{loc. max.}}$ ($U_{\mathrm{loc. max.}}$ is the local maximum of the potential between two wells) together with the corresponding wave functions $\y_{n}(\xi)$, are known. The pure tunneling dynamics occurs for packets of the following form\footnote{Formally, the wave functions $\y_n(\xi)$ and series coefficients $C_n$ are supposed to be complex. In fact, both of them are real for the under-barrier states. The whole wave function $\Phi(\xi,\t)$ is complex of course.}
\be
\Phi(\xi,\t)=\sum_{n} C_n e^{-i\bar{E}_n \t} \y_n(\xi),\quad \bar{E}_n< U_{\mathrm{loc. max.}},\quad \t=\w_0 t ,
\la{WP2levSer}
\ee
with
\be
C_n=\int \,dx\, \Phi(\xi,0) \y^*_n(\xi).
\la{Cn}
\ee
The probability $P^R(t)$ (or $P^R(\t)$ with the apparent modification of the result in below) to find a particle at the time $t$ in the definite local minimum $R$ is
\be
P^R(t)=\int_R \,dx \, |\Phi(x,t)|^2=\sum_{m,n} C^*_m C_n e^{i(E_m-E_n)t}\int_R\,dx\, \y^*_m(x) \y_n(x).
\la{PRdef}
\ee
Within the two-level approximation (when $n$ runs over $1$ and $2$ and the wave functions are real; cf. footnote 4),
\be
P^R(t)=P^R(0)-4 C_1 C_2 \sin^2 {(E_1-E_2)t} \int_R\,dx\,\y_1(x) \y_2(x).
\la{PR2lev}
\ee

\ssk
In the symmetric case (displayed in Fig.\ref{fig:image1}a), the wave packet \rf{wavepack} makes periodic oscillations between two minima, with the period $T=2\pi/\triangle$; here $\triangle=\bar{E}_{-1}-\bar{E}_{-2}=\n-\m$. We observe that the wave packet completely transfers from one well to another, so that the ``Josephson's oscillations''  \rf{PR2lev} take place. Having a deformation of $\tilde{U}(\xi,\bar{E}_{-2},\bar{E}_{-1},\L)$ from the symmetric shape, the picture of the tunneling dynamics is essentially modified (cf. Fig.\ref{fig:image2}). It is featured the so-called ``partial confinement'' of the wave packet in the initial well. The effect is amplified with growing the deformation of the potential (see Fig.\ref{fig:image2}b). Such a behavior of the tunneling dynamics is determined by a re-distribution of wave functions between local minima upon increasing the potential deformation. The wave packet recovery time coincides with that of the symmetric case because the value of the tunnel doublet does not change upon the potential deformation. Recall that the just discussed hallmarks of the tunneling dynamics take place within the two-level approximation for the potentials in Fig.\ref{fig:image1}, when the main contribution to the wave packet is given by the wave functions of the tunnel doublet.

\begin{figure}[h]
\begin{minipage}[h]{0.499\linewidth}
\center{\includegraphics[width=0.9\linewidth]{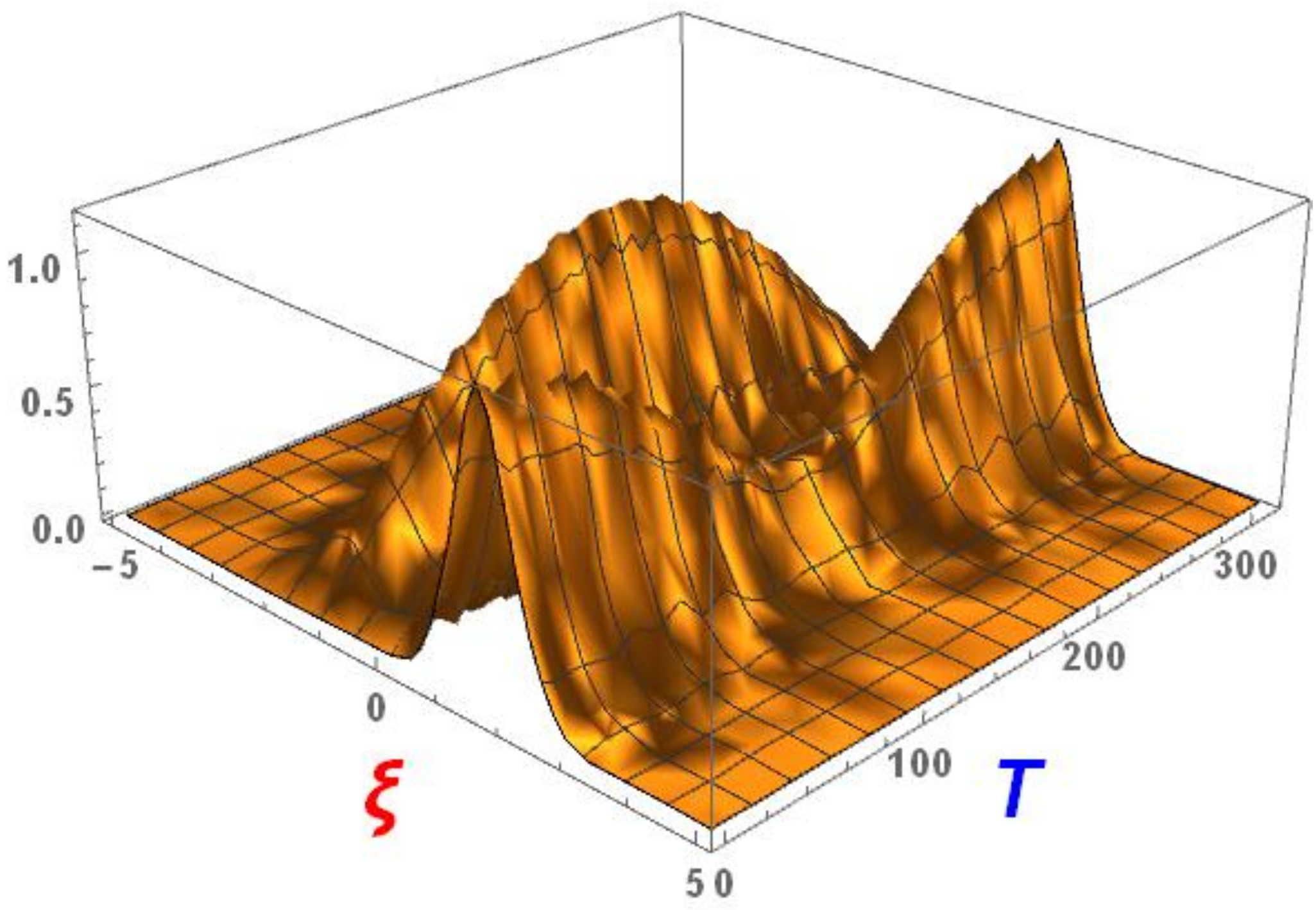} \\ a)}
\end{minipage}
\hspace{0.2cm}
\begin{minipage}[h]{0.499\linewidth}
\center{\includegraphics[width=1.\linewidth]{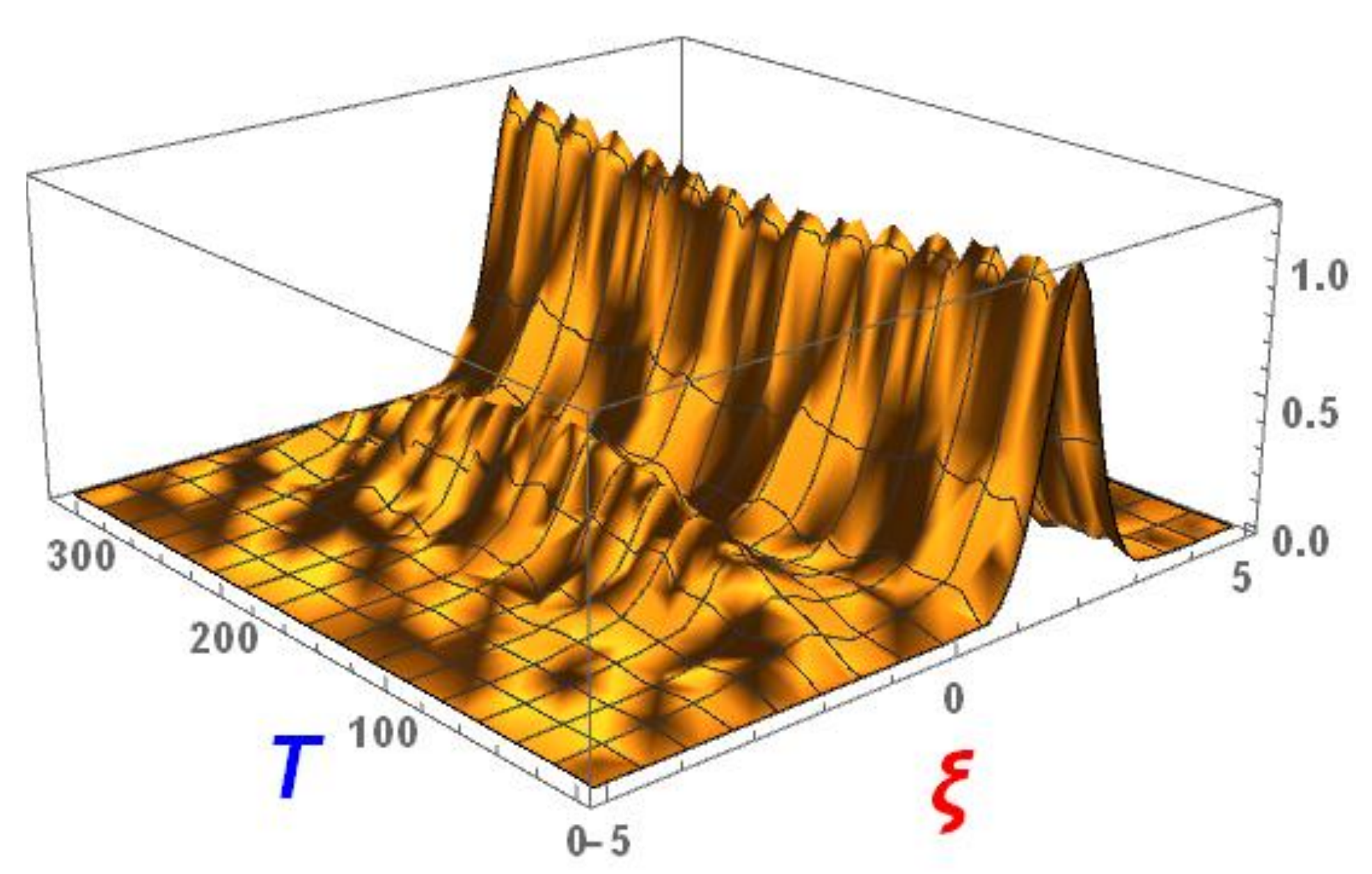} \\ b)}
\end{minipage}
\caption{$|\Phi(\xi,T)|$ in the case of $\tilde{H}$ with: a) $\n=-3.0$, $\m=-3.02$, $\Lambda=0.5$; b) $\n=-3.0$, $\m=-3.02$, $\Lambda=0.05$. $T$ and $\xi$ are changed in the following ranges: $-5\le \xi \le 5$, $0\le T \le 300$.}
\label{fig:image2}
\end{figure}

\ssk
To sum up, having the relevant basis in the disposal, it is always possible to present an arbitrary wave packet in its terms and, by use of the appropriate approximation, to get a qualitative characteristic of localization in the definite local minimum. Wave packets dynamics in two-well potentials $\tilde{U}(\xi,\bar{E}_{-2},\bar{E}_{-1},\L)$ within the two-level approximation will be characterized by values $P^R(\t)<1$, that means the absence of full localization in view of a non-trivial tunnel doublet value. The case of the complete localization of the tunnel doublet states corresponds to $P^R(\t)\equiv P^R(0)$.

\ssk
Dynamics of wave packets in three-well potentials within the taken here approach, ref. \cite{Berezovoj:2013}, have a number of differences from the previously considered case. Hamiltonian \rf{Htil} contains a set of parameters, in dependence on which the potential gets two or tree wells. When the first added level (of energy $\bar{E}_{-1}$ in our notation) is located near the ground state of the initial Hamiltonian $H_0$ and forms, with the $H_0$ ground state $\y_0$ (of energy $\bar{E}_{0}$),  the tunnel doublet of $(\y_{-1},\y_{0})$ states, the potential $\tilde{U}(\xi,\bar{E}_{-2},\bar{E}_{-1},\L)$ possesses three local minima (see Fig.\ref{fig:image3}).

\begin{figure}[h]
\hspace{-1.cm}
\begin{minipage}[h]{0.499\linewidth}
\center{\includegraphics[width=0.9\linewidth]{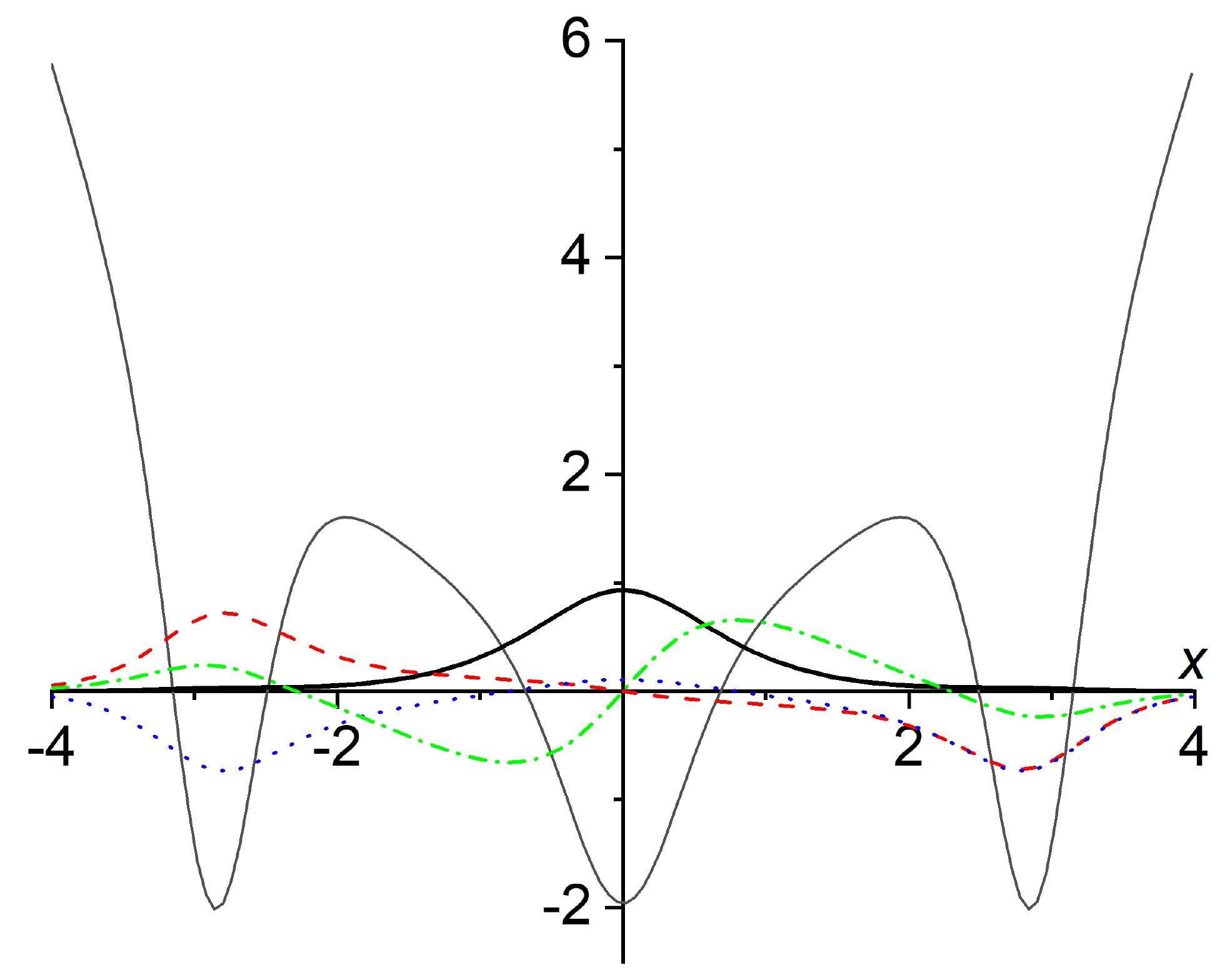} \\ a)}
\end{minipage}
\hspace{1.1cm}
\begin{minipage}[h]{0.499\linewidth}
\center{\includegraphics[width=1.\linewidth]{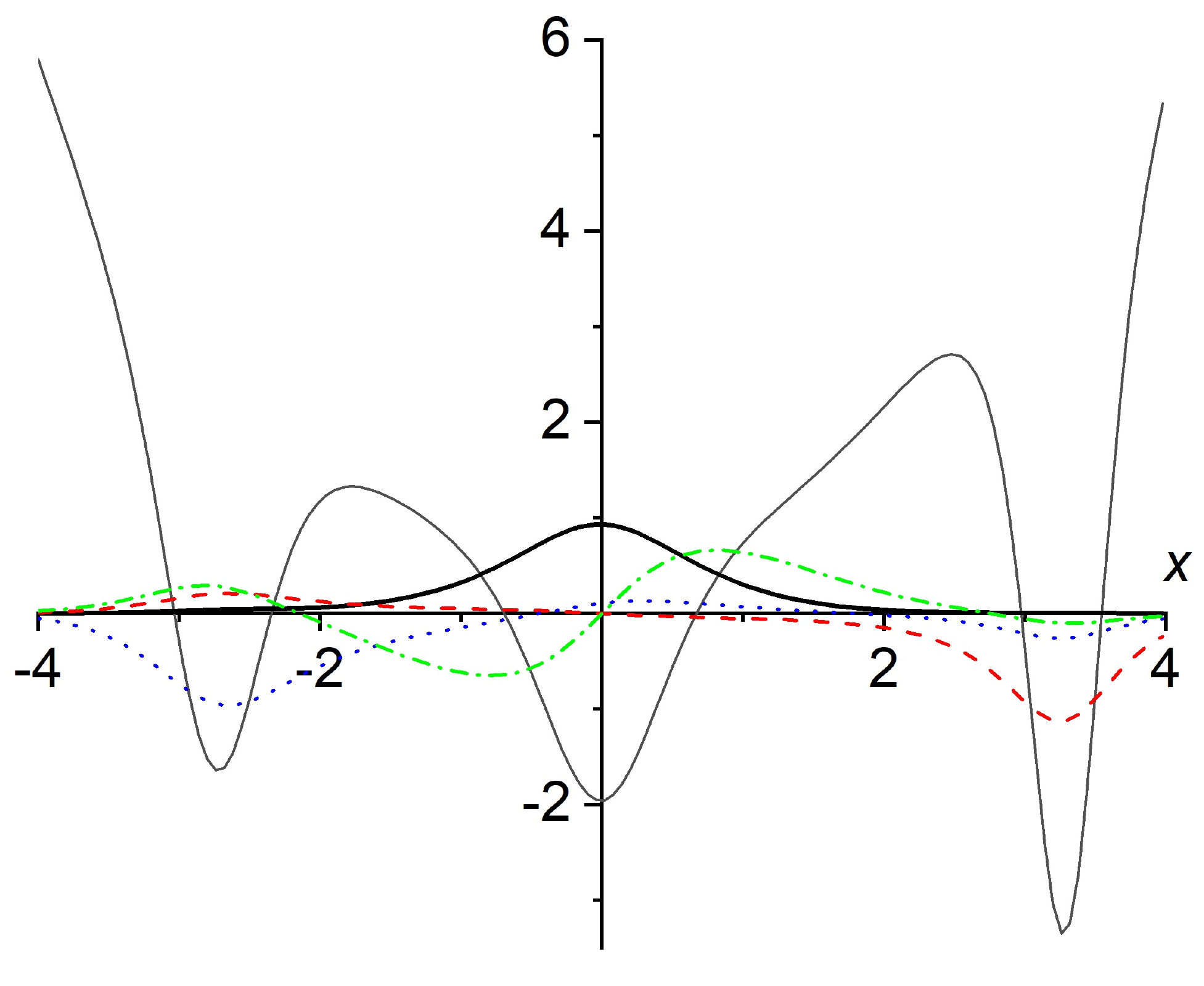} \\ b)}
\end{minipage}
\caption{The potential $\tilde{U}(\xi,\bar{E}_{-2},\bar{E}_{-1},\L)$ with three wells (thin solid lines) and wave functions of first levels. The parameters are: a) $\n=-0.02$, $\m=-1$, $\Lambda=1$; b) $\n=-0.02$, $\m=-1$, $\Lambda=0.05$. Bold solid lines correspond to the ground state wave functions; dashed lines are that of the first excited level, dotted lines correspond to the second excited level wave functions. (The dashed-dotted line is that of the third excited level.)}
\label{fig:image3}
\end{figure} 

\ssk
Figure \ref{fig:image3} shows that the ground state wave function of a (deformed) three-well potential is localized in the central minimum domain {\it independently} on values of the deformation parameter. The excited states are delocalized in between uttermost minima, and deforming the potential from the pure symmetric shape results in the redistribution of their wave functions over the uttermost wells. These facts impact the tunneling process of the wave packet, initially located in one of the utmost minima (see Fig.\ref{fig:image4}). Both cases in Fig.\ref{fig:image4} are characterized by a small value of $|\Phi|$ within the central minimum domain at any values of the deformation parameter. (We suppose the initial wave packet is in the right well.)

\begin{figure}[h]
\hspace{-1.3cm}
\begin{minipage}[h]{0.499\linewidth}
\center{\includegraphics[width=0.9\linewidth]{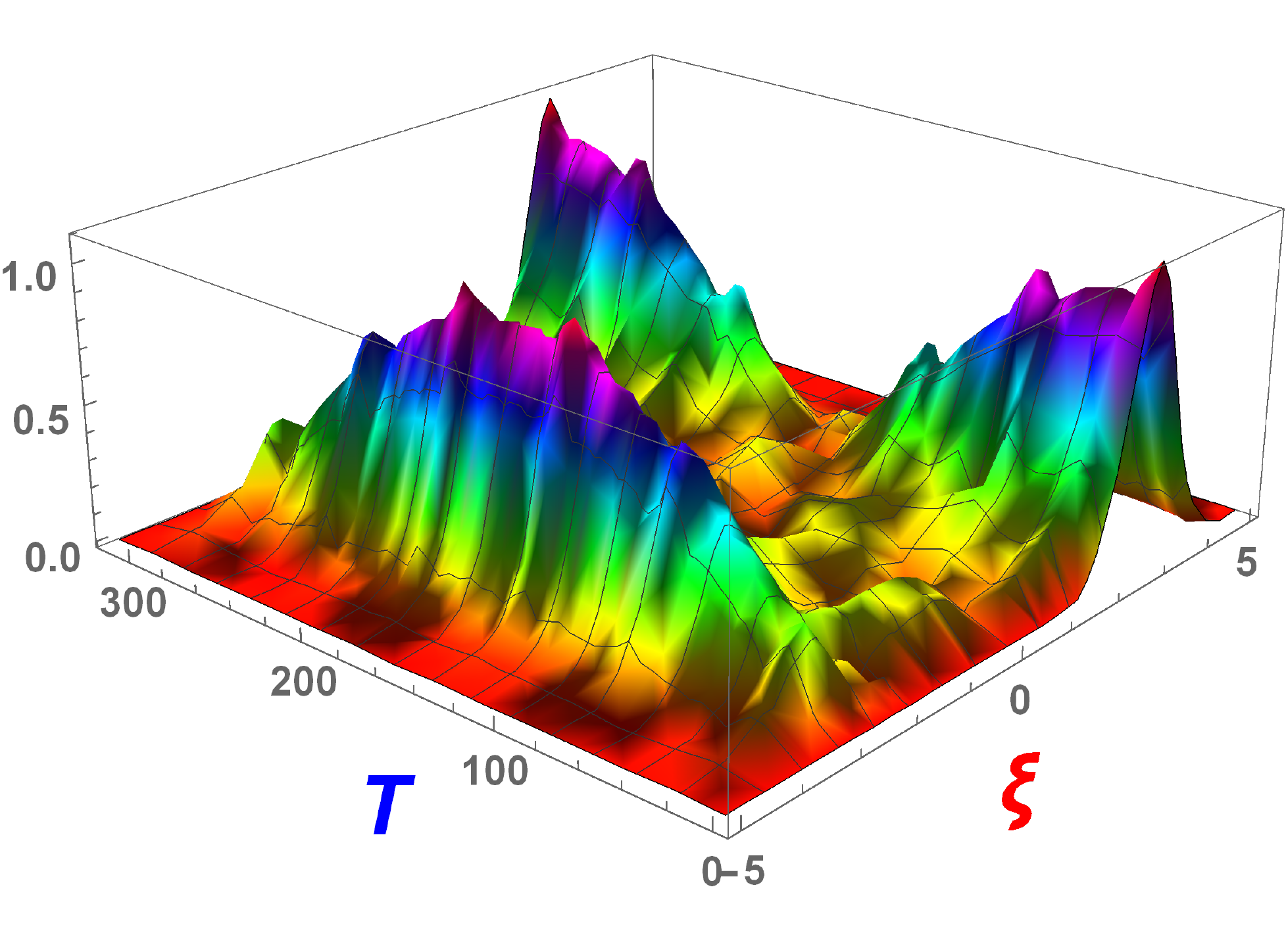} \\ a)}
\end{minipage}
\hspace{1.29cm}
\begin{minipage}[h]{0.499\linewidth}
\center{\includegraphics[width=0.98\linewidth]{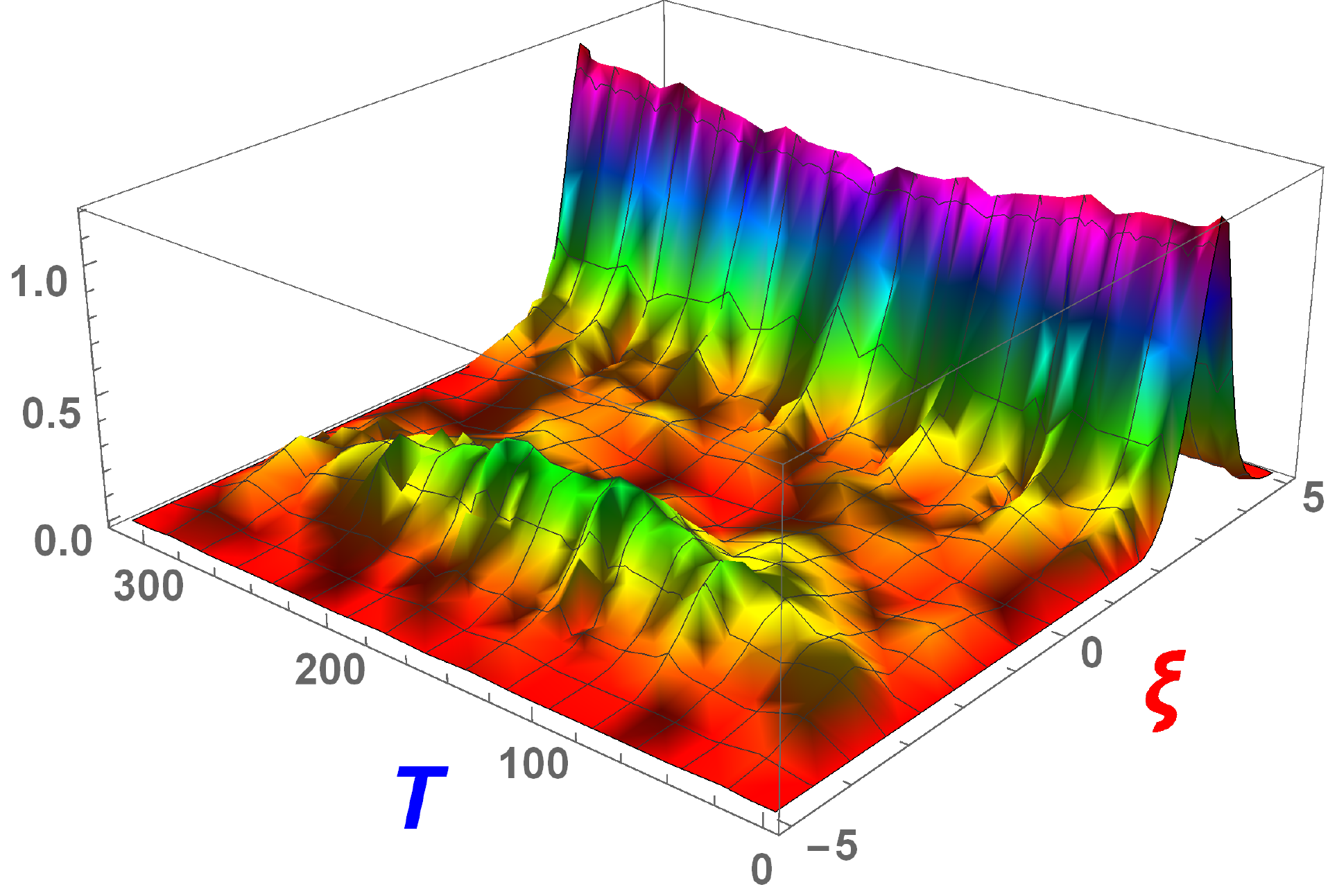} \\ b)}
\end{minipage}
\caption{Two plots of $|\Phi(\xi,\t)|$ for the parameters of $\tilde{U}(\xi,\bar{E}_{-2},\bar{E}_{-1},\L)$ in Fig.3: a) $\n=-0.02$, $\m=-1$, $\Lambda=1$; b) $\n=-0.02$, $\m=-1$, $\Lambda=0.05$. The ranges of $\t\equiv T$ and $\xi$ are chosen to be: $-5\le \xi \le 5$, $0\le T \le 300$.
}
\label{fig:image4}
\end{figure}

\section{Localization of under-barrier states in asymmetric multi-well potentials}

Studies of the wave packets dynamics in an asymmetric double well, ref. \cite{Nieto:1985}, showed the possibility of the complete localization of quantum states in the false vacuum (the local minimum of the higher energy), in dependence on the deformation degree. Furthermore, in ref. \cite{Graffi:1984}, it was shown that the tunneling dynamics of symmetric double-well Hamiltonians is exponentially sensitive to a tiny perturbation of the potential, that results in localization of the under-barrier states. The disturbance potential is localized in one of the wells, far from the corresponding minimum; its value is chosen to be $|V_1|\gg \triangle$ ($\triangle$ is the tunnel doublet value). The sign of the potential $V_1$ determines the well, in which the under-barrier states will be localized. In ref. \cite{Landsman:2013}, localization of states was demonstrated by use of a tricky procedure of the direct integration of the Schr\"odinger equation with $U=V_0(x^2-a^2)^2$ potential and with\footnote{For the sake of generality, we will consider the $(x,t)$ set of the natural variables in analytical expressions of this section. Numerical computations are performed in terms of $(\xi,\t)$.} 
\be
V_1(x)=\left\{
\begin{array}{l}
s\cdot \exp\left(\fr1{c^2}-\fr{1}{c^2-(x-b)^2}\right) \qquad \mathrm{if}\,\,|x-b|<c \\
0 \qquad \qquad \qquad \qquad \qquad \quad\,\, \mathrm{if}\,\,|x-b|>c \,\,.
\end{array}
\right.
\la{V1def}
\ee

\ssk
Below we will show the advantages of using the basis \rf{Psitil0}-\rf{Psitili} in studying localization of states of symmetric Hamiltonians with two- and three-well potentials, on account of the disturbance \rf{V1def}. The spectrum and the wave functions of the disturbed Hamiltonian $\hat{H}=\tilde{H}+V_1$ (recall, $\tilde{H}$ is Hamiltonian \rf{Htil} with the HO Hamiltonian $H_0$) are recovered from the diagonalization procedure for $\hat{H}\Psi^{(n)}(x)=\hat{E}_n\Psi^{(n)}(x)$, with the general expansion of wave functions:
\[
\Psi^{(n)}(x)=\tilde{C}^{(n)}_0 \tilde{\Psi}_0(x,E_{-2},\L_{-2};E_{-1},\L_{-1})+\tilde{C}^{(n)}_1 \tilde{\Psi}_1(x,E_{-2},\L_{-2};E_{-1},\L_{-1})
\]
\be
+\sum^N_{i=0}\tilde{C}^{(n)}_{i+2} \tilde{\Psi}_{i+2}(x,E_i),\qquad n=0,1,\dots,N \, ,
\la{Psidef}
\ee
with $\Lambda_{-2}=1=\Lambda_{-1}$.
Here $N$ is a number, from which the spectrum $\hat{E}_n$, $n>N$ and the wave functions $\tilde{\Psi}^{(n)}(x)$ are exactly that of the Hamiltonian $\tilde{H}$ (i.e., the wave functions of eqs. \rf{Psitil0}-\rf{Psitili}).\footnote{For the parameters $s$, $b$ and $c$ of eq. \rf{V1def}, which are used in our subsequent calculations, we established, by the trial and error method, $N=7$.}

\ssk
After diagonalizing the disturbed Hamiltonian $\hat{H}$, we can compute the probability density of the corresponding wave functions and figure out the localization degree of the lowest states, forming the tunnel doublet (see Fig.\ref{fig:image5}). From the shapes of $|\Psi^{(0)}|^2$ and $|\Psi^{(1)}|^2$ in Fig.\ref{fig:image5}, it follows that, upon increasing the disturbance parameters, the degree of localization growths and becomes full at $s=1$. 
The energy spectrum of first ten states of the disturbed Hamiltonian is given in Table 1.


\begin{figure}[h!]
\hspace{-1.cm}
\begin{minipage}[h]{0.48\linewidth}
\center{\includegraphics[width=0.9\linewidth]{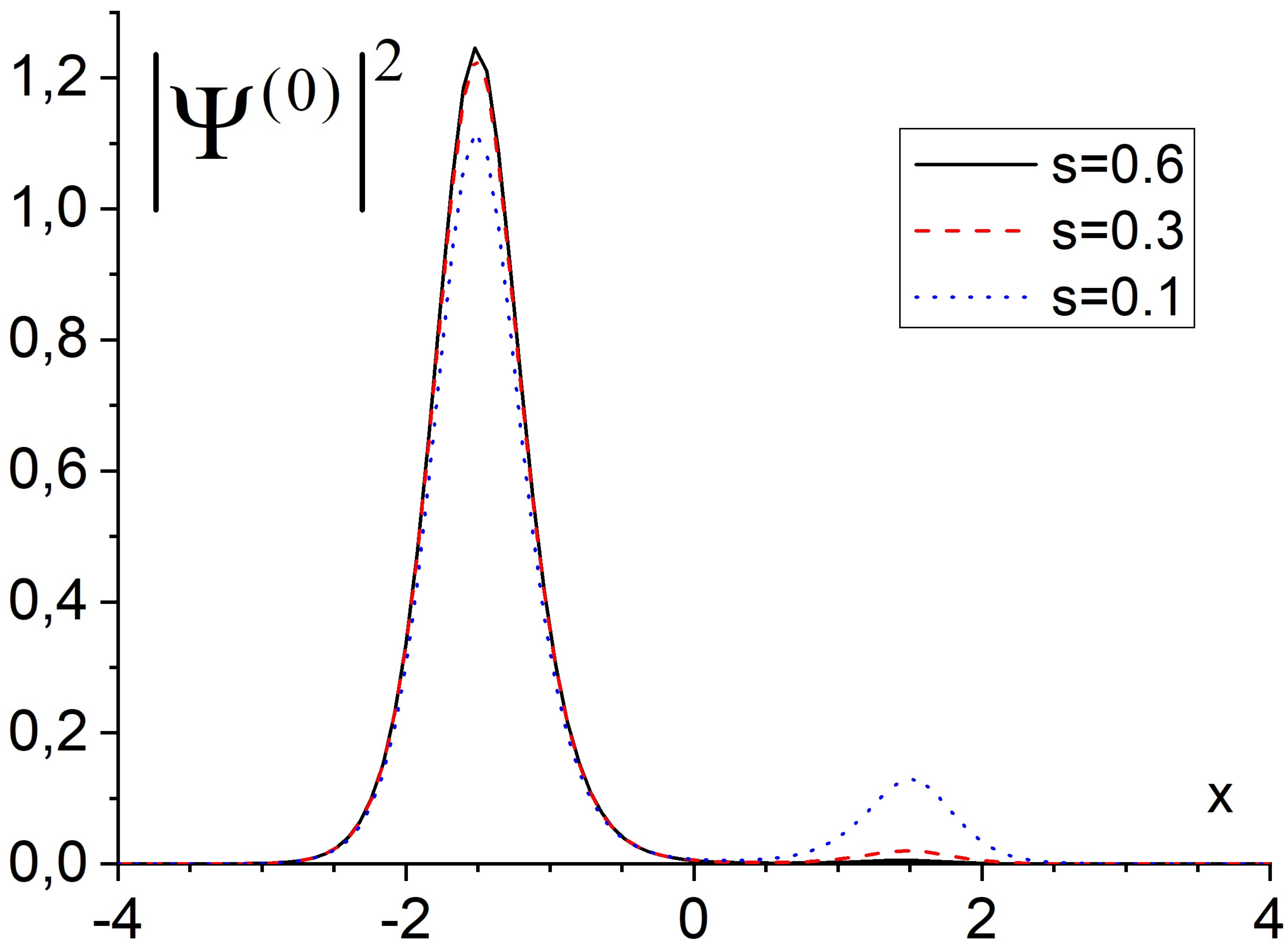} }
\end{minipage}
\hfill
\begin{minipage}[h]{0.48\linewidth}
\center{\includegraphics[width=0.9\linewidth]{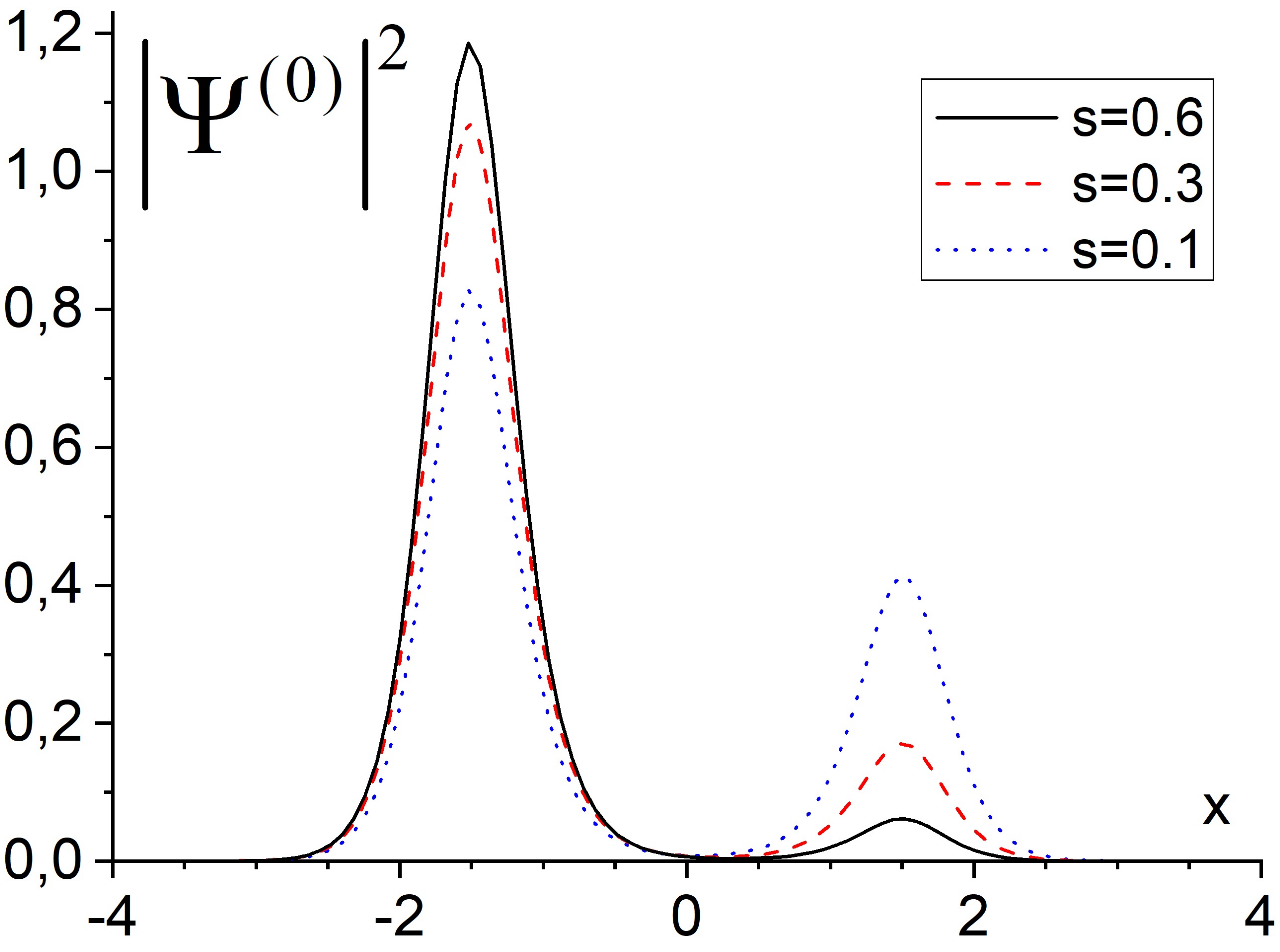} }
\end{minipage}
\hfill
\centerline{a)}
\hfill
\begin{minipage}[h]{0.48\linewidth}
\center{\includegraphics[width=0.9\linewidth]{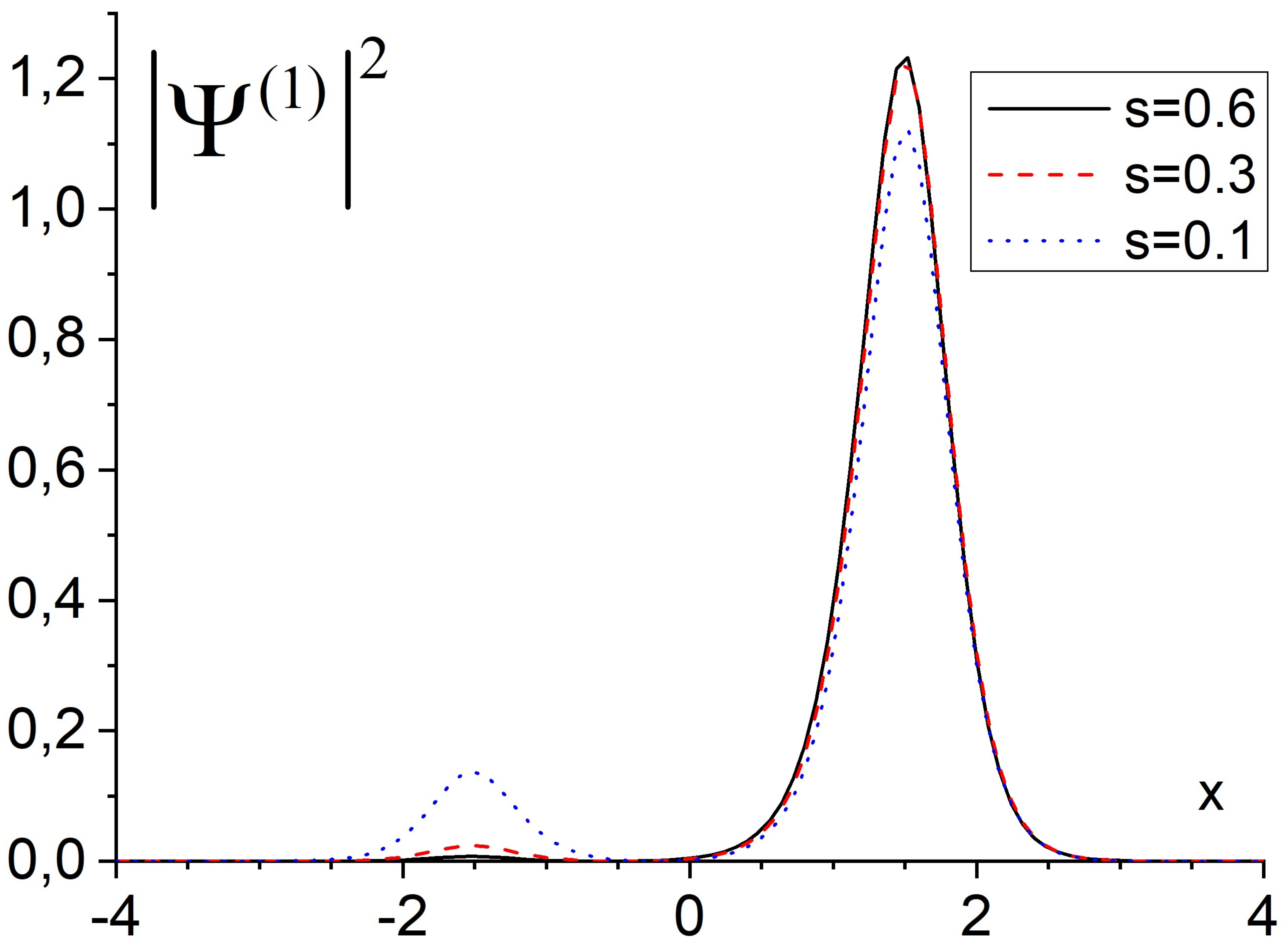} }
\end{minipage}
\hfill
\hspace{0.99cm}
\begin{minipage}[h]{0.499\linewidth}
\center{\includegraphics[width=0.9\linewidth]{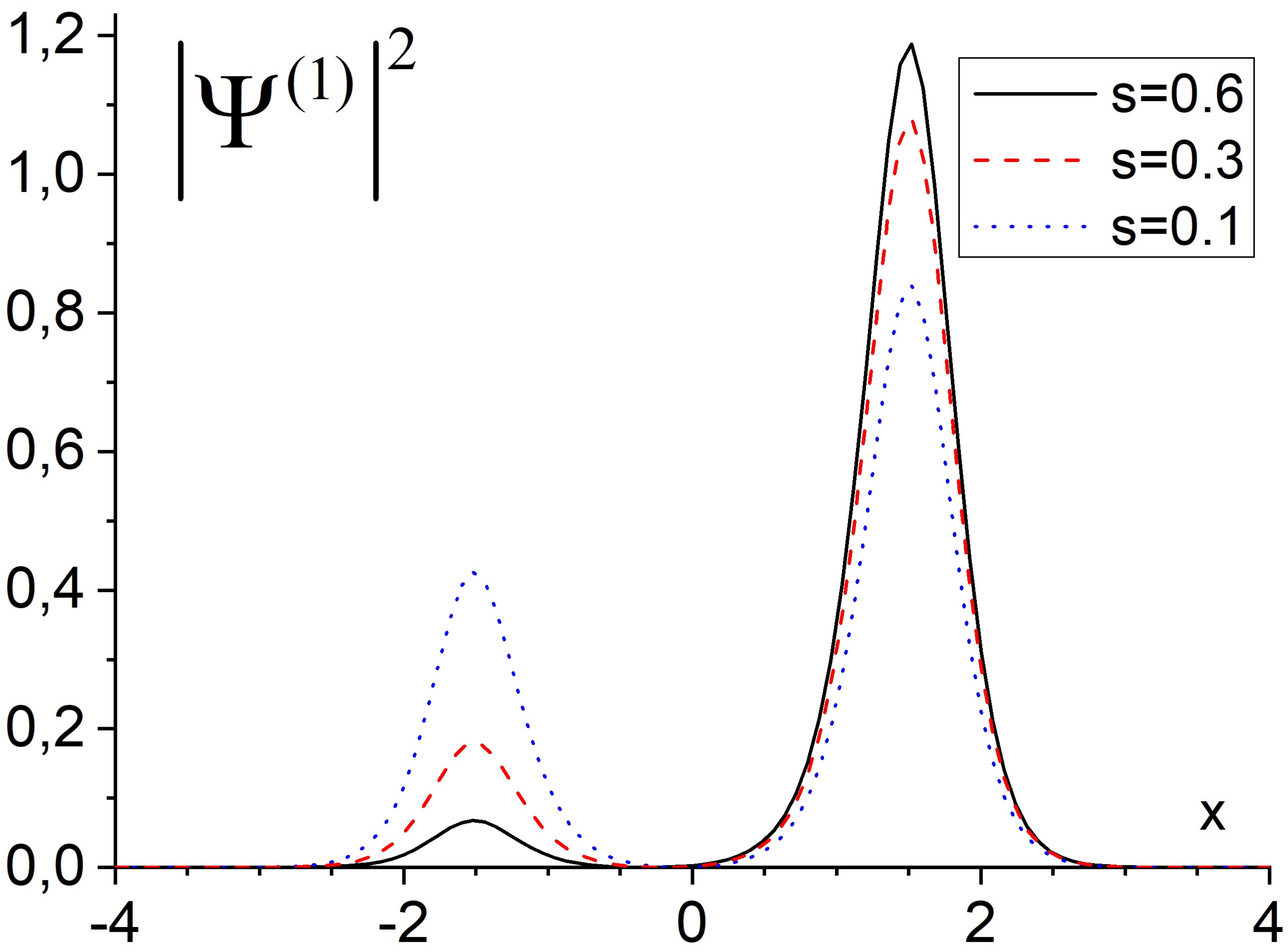} }
\end{minipage}
\centerline{b)}
\caption{a) Localization of the ground state wave function of the disturbed Hamiltonian with the disturbance parameters $b=1.86$, $c=0.5$ (left panel) and $b=1.86$, $c=0.25$ (right panel);
b) Localization of the first excited state wave function of the disturbed Hamiltonian with the disturbance parameters $b=1.86$, $c=0.5$ (left panel) and $b=1.86$, $c=0.25$ (right panel). Other parameters of the Hamiltonian are the same as in Fig.1a.}
\label{fig:image5}
\end{figure}


\begin{table}[ht]
\small
\hspace{-0.4cm}
\begin{tabular}{c| c c c c c c c c c c} 
\hline\hline 
Parameters & $n$=0 & $n$=1 & $n$=2 & $n$=3 & $n$=4 & $n$=5 & $n$=6 & $n$=7 & $n$=8 & $n$=9 \\ [0.5ex] 
\hline 
\\ 
$b$=1.86, $c$=0.25, $s$=0.6 & -2.512 & -2.466 & 0.507 & 1.515 &2.506 & 3.501 & 4.509 & 5.509 & 6.501 & 7.509  \\ 
[0.5ex]  \hline\hline \\
$b$=1.86, $c$=0.5, $s$=0.6 & -2.511 & -2.357 & 0.525 & 1.549 & 2.526 & 3.508 & 4.529 & 5.521 & 6.510 & 7.524 \\ [1ex] 
\hline \hline 
\end{tabular}
\label{table:energy} 
\caption{The energy spectrum of the distrubed Hamiltonian: first ten levels.}
\end{table}

\ssk
Looking at Table 1, one may notice that the chosen disturbance does not lead to essential changes in the HO energies $1/2,3/2,5/2,\dots$ upper the tunnel doublet. On the other hand, the value of the tunnel doublet by itself (recall, it has been originally chosen to be $\triangle=0.02$) gets essentially increased: $\triangle=0.046$  for $c=0.25$ and $\triangle=0.154$ for $c=0.5$. It decisively matters in localization of states of a Hamiltonian with two symmetric wells, upon incorporating the interaction, which breaks the reflective symmetry.

\ssk
Let us present the values of first coefficients in the expansion \rf{Psidef} for the ground and first excited level wave functions, Table 2. Looking on them, one may notify the essential contribution into the under-barrier wave function of the disturbed Hamiltonian comes from the states of the tunnel doublet of $\tilde{H}$. The contribution of more higher states of $\tilde{H}$ is small, though their values are comparable to each other. It mainly occurs due to the choice of parameters of $\tilde{H}$, that provides a good enough fulfillment of the two-level approximation. Graphically, the choice of the parameters for the two-level approximation becomes apparent (see Fig.\ref{fig:image6-1}a and Fig.\ref{fig:image6-1}b): they do not deform the initial potential much. However, the Schr\"odinger equation $\hat{H} \Psi^{(n)}=\hat{E}_n \Psi^{(n)}$ can not only be satisfied due to the tunnel doublet wave functions. Despite the small fraction of states above the tunnel doublet in the wave function \rf{Psidef}, their contribution to the formation of the full potential (see the discussion around \rf{UErel} below) is significant: the contribution of the under-barrier states within the disturbance region is commonly small, so the uttermost contribution comes from the high-level states. In the case of an analysis of the localization problem in a simplest model of two states \cite{Bolotin:1993}, the contribution of higher states is phenomenologically taken into account by the asymmetry degree parameter.

\ssk
\begin{table}[ht]
\small
\centering 
\begin{tabular}{c| c| c c c c c } 
\hline\hline  
&&&&&&\\
Parameters & Wave functions & $\tilde{C}^{(0)}_0$ & $\tilde{C}^{(0)}_1$ & $\tilde{C}^{(0)}_2$ & $\tilde{C}^{(0)}_3$ & $\tilde{C}^{(0)}_4$  \\ [0.5ex] 
&  & $\tilde{C}^{(1)}_0$ & $\tilde{C}^{(1)}_1$ & $\tilde{C}^{(1)}_2$ & $\tilde{C}^{(1)}_3$ & $\tilde{C}^{(1)}_4$  \\ [0.5ex] 
\hline 
\\ 
$b$=1.86, $c$=0.25, $s$=0.6 & $\Psi^{(0)}$ & 0.8491 & 0.5283 & 0.0013 & 0.0013 & -0.0007   \\ 
 & $\Psi^{(1)}$ & 0.5286 & -0.849 & 0.0059 & 0.006 & 0.003\\
[0.5ex]  \hline\hline \\
$b$=1.86, $c$=0.5, $s$=0.6 & $\Psi^{(0)}$ & 0.7558 & 0.6548 & 0.0013 & 0.0012 & 0.0005   \\ 
 & $\Psi^{(1)}$ & 0.6544 & -0.7555 & 0.0205 & 0.0197 & 0.0087 \\[1ex] 
\hline \hline 
\end{tabular}
\label{table:wf} 
\caption{The contribution of states of the non-disturbed Hamiltonian $\tilde{H}$ into the lowest wave functions of $\hat{H}$.}
\end{table}

\ssk
As a criterium to verify the precision of our computations, we used a relation that links the full potential (i.e., the potential of $\hat{H}$) to the computed, within the diagonalization procedure, ground state wave functions and ground state energies. It is known, ref. \cite{Plyushchay:2017}, that these values are related to each other via
\be
\hat{U}(x)-\hat{E}_0=-\fr{\hbar^2}{4m} S(\hat{f}(x)),
\la{UErel}
\ee
where $S(f(x))$ denotes the Schwarzian derivative 
\be
S(f(x))=\left(\fr{f^{''}}{f'} \right)-\fr12 \left(\fr{f^{''}}{f'} \right)^2,
\la{Schwarzian}
\ee
for the function 
\be
{ \hat{f}(x)=-\int_x \,ds\, \fr1{\Psi_{(0)}^2(s)} }.
\la{fpsi0}
\ee
I.e.,
\be
{ S(\hat{f}(x))=-2\left(\fr{d^2}{dx^2}\ln \Psi^{(0)}(x) +\left(\fr{d}{dx}\ln \Psi^{(0)}(x) \right)^2 \right) .}
\la{Spsi0}
\ee
By use of eq. \rf{UErel} with the r.h.s. of eq. \rf{Spsi0}, we find that the results of Tables 1 and 2 are in agreement to the disturbance potential \rf{V1def} with the parameters, indicated inside the Tables.\footnote{Note that a similar to eq. \rf{UErel} relation can also be found for excited levels of the complete Hamiltonian $\hat{H}$, though it will require additional steps to isolate zeros in the corresponding wave functions.} Since the ground state function of the perturbed Hamiltonian is formed by a finite set of the HO wave functions (see eq. \rf{Psidef}), we conclude that our choice on the number of the basic wave functions (recall, $N = 7$ in our case) works fine. Another way to justify the chosen cut-off in the series expansion \rf{Psidef} is to compare the results of the diagonalization for higher values of the cut-off with that of Tables 1 and 2. One may verify that, with higher values of $N$ (we checked it up to $N=17$), the energy spectrum and the series expansion coefficients for the first ten levels are almost matched with values of Tables 1 and 2; for levels higher than ten, the spectrum and the wave functions are not visibly changed with adding new levels.

\begin{figure}[h!]
\hspace{-1.cm}
\begin{minipage}[h]{0.48\linewidth}
\center{\includegraphics[width=1.45\linewidth]{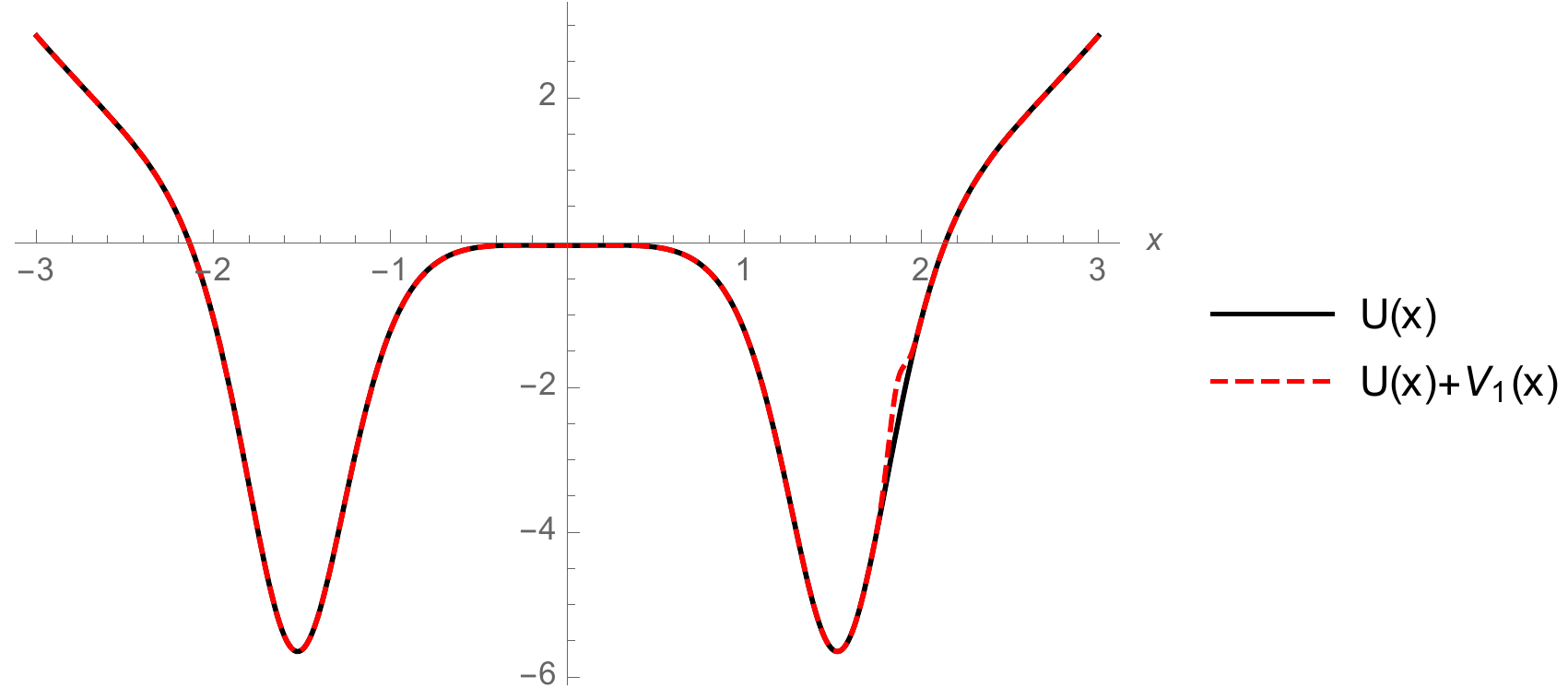} \\ a)}
\end{minipage}
\hfill
\begin{minipage}[h]{0.48\linewidth}
\center{\includegraphics[width=1.1\linewidth]{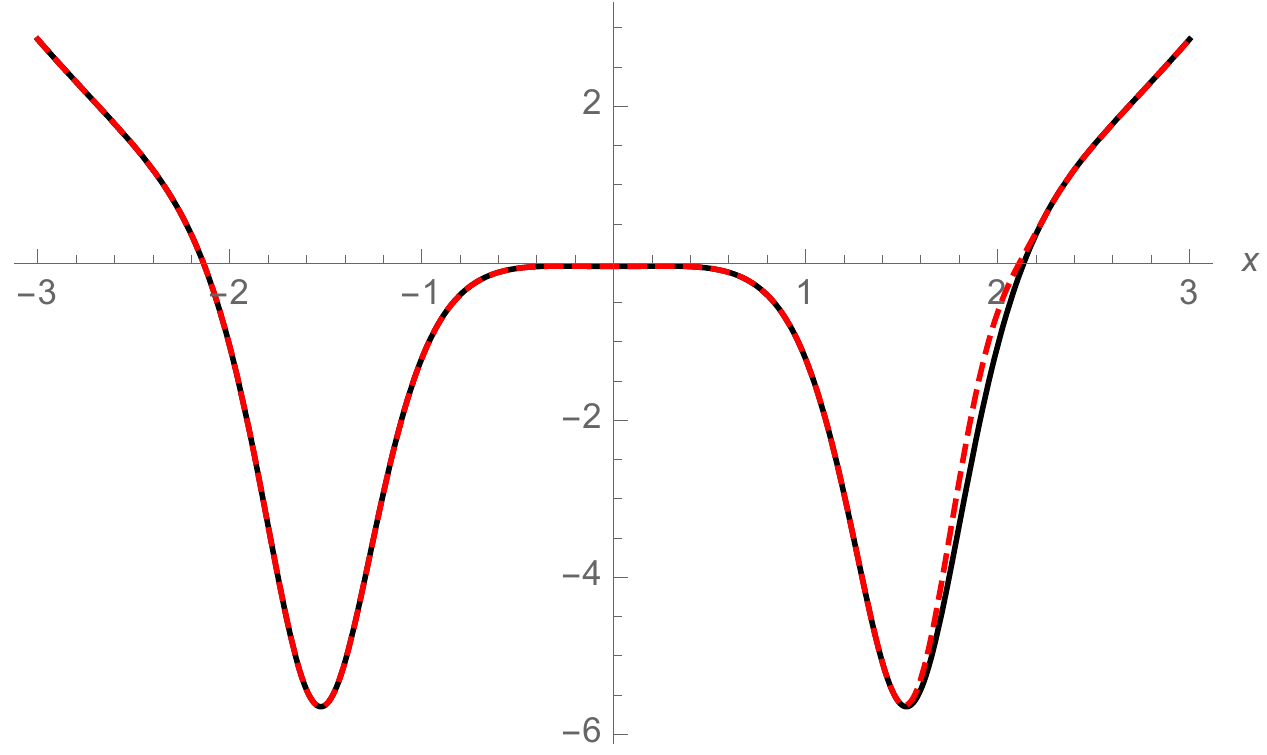} \\ b)}
\end{minipage}
\hfill
\begin{minipage}[h]{0.48\linewidth}
\center{\includegraphics[width=1.25\linewidth]{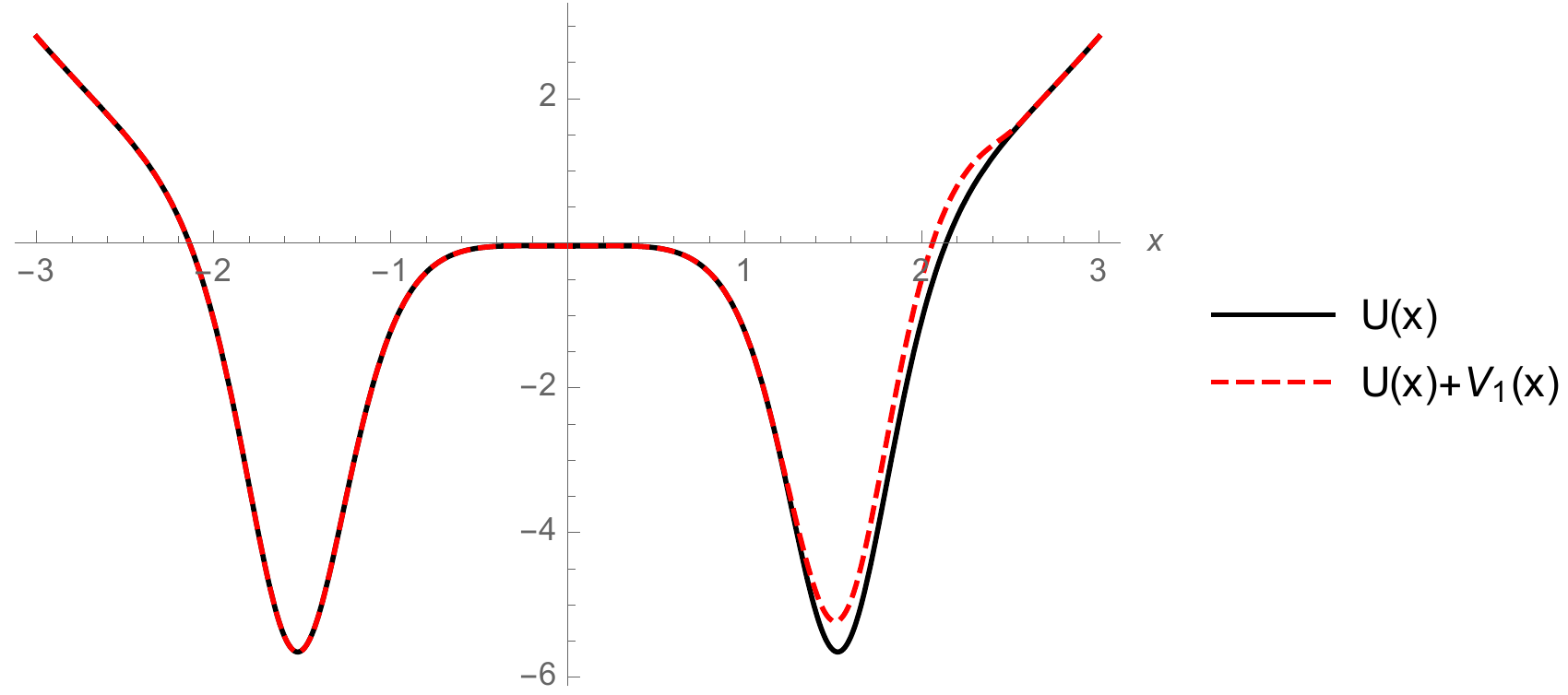} \\ c)}
\end{minipage}
\hfill
\hspace{0.99cm}
\begin{minipage}[h]{0.499\linewidth}
\center{\includegraphics[width=1.0\linewidth]{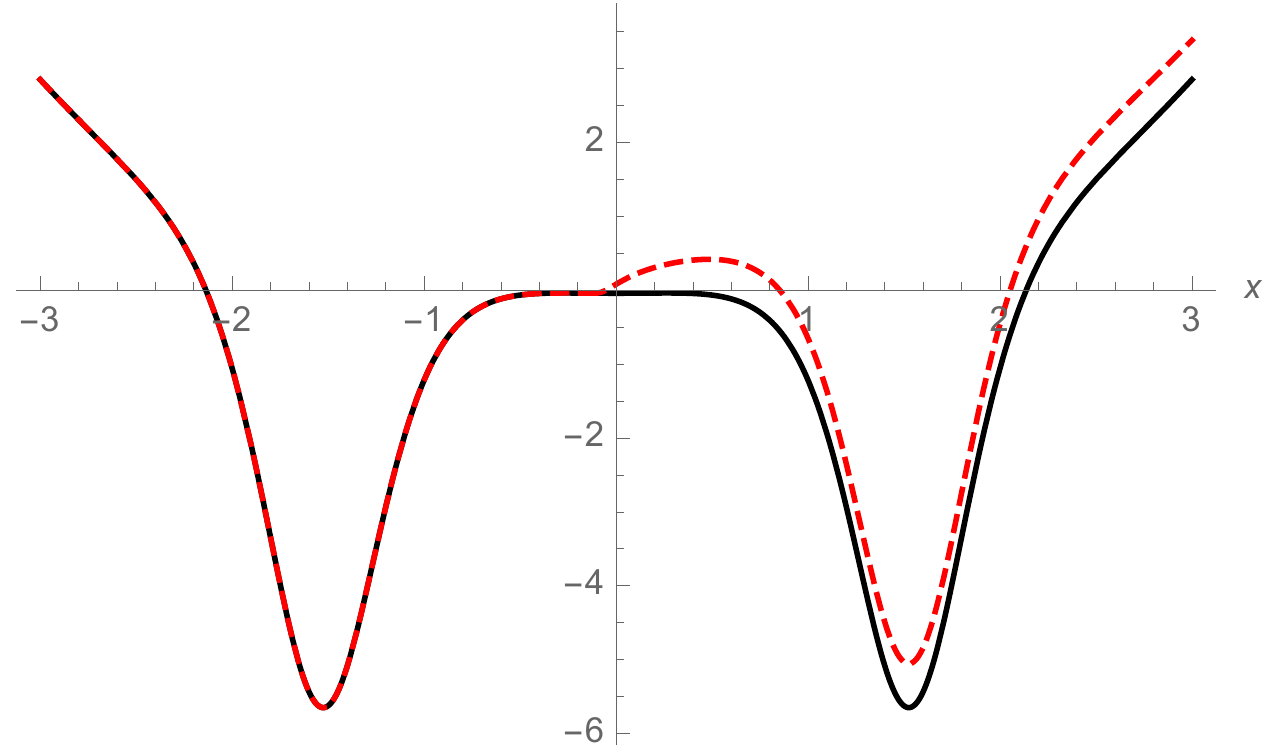} \\ d)}
\end{minipage}
\caption{The disturbance of a two-well symmetric potential in Fig.1a with $V_1(x)$ of eq. \rf{V1def} with $s=0.6$ and $b=1.86$. Other parameters are choosen to be: a) $c=0.25$; b) $c=0.5$; c) $c=0.8$; d) $c=2.0$. The upper panel shows the disturbance, to which the two-level approximation is applicable. }
\label{fig:image6-1}
\end{figure} 

\ssk
Thus, we have validated that localization of the under-barrier states is realized on a small number of the basic wave functions upon the relevant choice of the basis. In the case, when the (small) disturbance of the main potential is located far from the local minima of the, by now, full potential (like in Fig.\ref{fig:image6-1}c), one needs to take into account more basic states in the series expansion for the wave functions (akin to eq. \rf{Psidef}). Indeed, forming the spectrum and the wave functions of the disturbed Hamiltonian is strongly determined by the over-barrier states, hence we need to increase the number of the basic wave functions. (Our computations shows that the full potential can be effectively recovered on account of about twenty basic states.)

\ssk

Localization of states takes place in the vicinity of the quasi-crossing of energy levels of undisturbed Hamiltonian, and the localization degree depends on the value of the potential disturbance. In the case of a three-well potential, the ground state is localized in the central well, whereas the first and the second excited levels are delocalized (cf. Fig.\ref{fig:image3}). Taking into account the disturbance \rf{V1def} in the far right well of a symmetric potential at a special choice of its parameters, one may reach the complete localization of states. (See Fig.\ref{fig:image7}.)
\begin{figure}[h!]
\hspace{-1.cm}
\begin{minipage}[h]{0.48\linewidth}
\center{\includegraphics[width=0.9\linewidth]{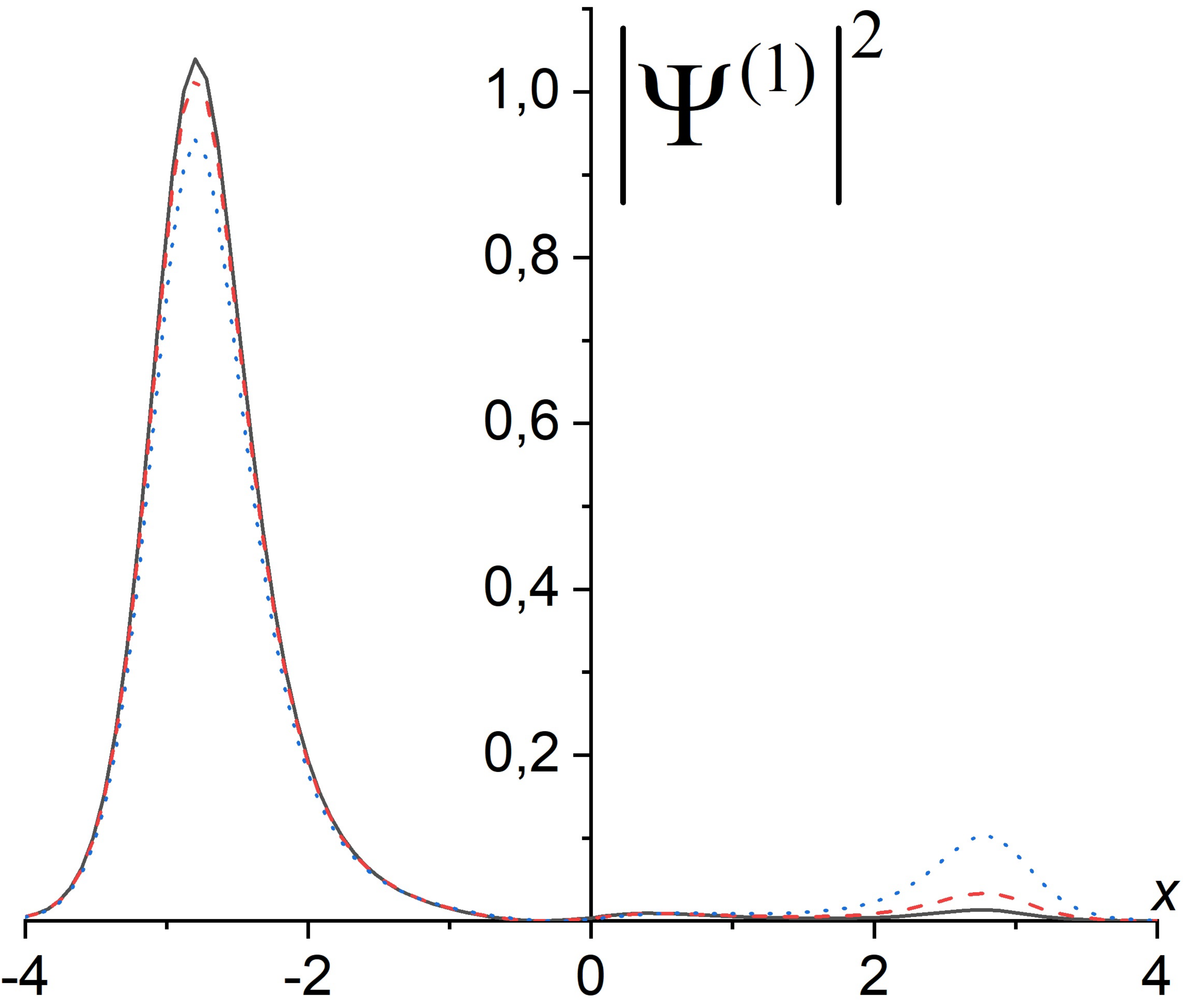} \\ a)}
\end{minipage}
\hfill
\begin{minipage}[h]{0.48\linewidth}
\center{\includegraphics[width=0.9\linewidth]{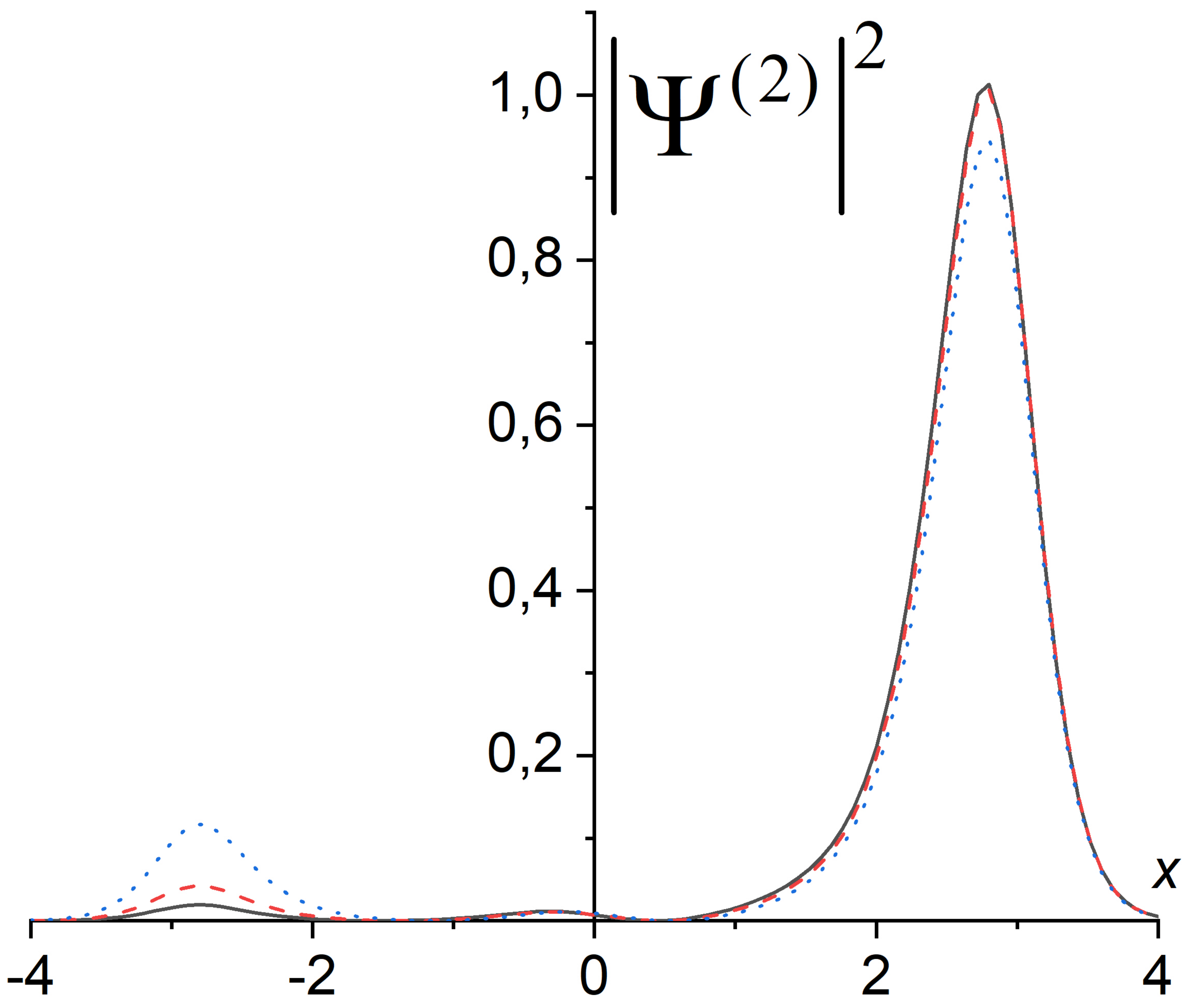} \\ b)}
\end{minipage}
\caption{Localization of the first and the second excited states of the symmetric three-well potential ($c=0.25$, $b=3$). }
\label{fig:image7}
\end{figure} 
One may notice non-zero values of the wave functions outside of the main peaks, that is a reflection of having the nodes in the corresponding wave functions.

\ssk

We end up this section with emphasizing that localization of initially delocalized states in symmetric potentials is accompanied with essential increasing the tunnel doublet value. When the perturbation, which breaks the reflective symmetry, is turned on,  the lion-share of the localized state wave functions consists of the contribution of the over-barrier states.

\section{Coherent control with an external driving force }

Interesting possibilities in constructing a new type of quantum devices open upon employing the control by external time-dependent fields on quantum-mechanical processes with keeping the coherence. Theoretical and experimental studies of quantum mechanical systems with external driving forces are very popular and widespread in view of their importance for modeling q-bits, the quantum telecommunication and transportation, developing new schemes of encryption and decryption, creation of new quantum interferometers and other tools for the highly precision measurement. Clearly, tunneling processes are a significant part of this activity.

\ssk
From a variety of the observed in the tunneling processes effects, it is worth mentioning theoretical predictions of the coherent tunneling destruction (CTD) \cite{Grossmann:1991,Longhi:2005} in symmetric two-well potentials, and experimental observations of this phenomenon \cite{Kierig:2008,DellaValle:2007}. Importance of the CTD consists in the possibility to control the tunneling dynamics of wave packets, up to their complete stop, by changing parameters of an external time-dependent field. We will reveal further on features of the tunneling in a system with general Hamiltonian \rf{Htil}, whose potentials are of a various deformation degree, in the presence of a periodic time-dependent perturbation.

\ssk
The evolution of quantum systems, characterized by periodic in time Hamiltonians $H(t)=H(t+T)$, is traditionally described in the Floquet wave functions basis:
\be
\Psi_\a (x,t)=\sum_\a c_\a \exp\left(-{i \ve_\a t} \right) \Phi_\a(x,t) ,\quad \Phi_\a (x,t)=\Phi_\a(x,t+T),
\la{Floquetpsi}
\ee
where $c_\a$ is determined by initial conditions, while the quasi-energies $\ve_\a$ and the Floquet wave functions $\Phi_\a (x,t)$ follow from the diagonalization procedure of the complete Hamiltonian.\footnote{Under the diagonalization we mean any of the methods to determine the quasi-energies and wave functions. They, in particular, include the Floquet matrix and matrix-continued-fraction methods (see, e.g., \cite{Grossmann:1991ZP,Dittrich:1998Book}).} The time periodicity of the Floquet wave functions makes possible to use the following expansion over the Fourier modes,
\be
\Phi_\a(x, t)=\sum_{n=-\infty}^{n=\infty} c^n_\a(x) e^{in\w t},
\la{FloquetFourier}
\ee
with the subsequent expansion of $c^n_\a(x)$ over the chosen complete orthonormal basis $\{\vf_k(x)\}_{k=1}^\infty$, so that
\be
\Phi_\a(x, t)=\sum_{k=1}^\infty \sum_{n=-\infty}^{n=\infty} c^{n}_{\a,k}\,\vf_k(x) e^{i n\w t} .
\la{Fourier}
\ee

\ssk
The notable example of time-periodic Hamiltonians is given by a symmetric two-well potential with an external time-periodic driving force \cite{Grossmann:1991,Grossmann:1991ZP}.
Following ref. \cite{Grossmann:1991ZP}, one finds the energy spectrum and the Floquet wave functions from the diagonalization of the following Hamiltonian
\be
\hat{H}(\xi,p_\xi,\t)=\fr{p^2}{2}+\tilde{U}(\xi,\bar{E}_{-2},\bar{E}_{-1},\L)+\xi S \cos(w \t+\vf_0),\quad \t=\w_0 t,
\la{Ht0}
\ee
where $w=\w/\w_0$ is the dimensionless ratio of the driving force frequency $\w$ to the internal frequency of the HO. In what follows, we choose the phase $\vf_0$ to be equal to zero.

\ssk
In contrast to the approach of ref. \cite{Grossmann:1991ZP}, where the basis $\{\vf_k(x)\}_{k=1}^\infty$ consists of the wave functions of the unperturbed harmonic oscillator, we use wave functions of an exactly-solvable Hamiltonian with symmetric and asymmetric potentials (cf. eqs. \rf{Psitil0}-\rf{Psitili}). The employment of an exactly-solvable model allows us to consider cases with the predefined deformation of the potential and the computed analytically spectrum of the under-barrier states, as well as to crucially decrease the dimension of a diagonalized matrix upon computing the energies and the Floquet wave functions.\footnote{For instance, the computations of ref. \cite{Llorente:1992} were performed on the basis set of 200 states of a harmonic oscillator with the optimized frequency. We use 10 basic states of the unperturbed exactly-solvable Hamiltonian instead.} Recall that the value of the initial tunnel doublet does not depend on the deformation degree within the considered approach. That, in particular, means changing the tunnel doublet value is completely determined by the external time-dependent driving force. States of the Hamiltonian $H(\xi,p_\xi)=p^2/2+\tilde{U}(\xi,\bar{E}_{-2},\bar{E}_{-1},\L)$ are suitable for basic functions of the Higgs type phenomenological potential as soon as the minima locations and the depth of wells of symmetric potentials will be reconciled by use of the parametric dependence of $\tilde{U}(\xi,\bar{E}_{-2},\bar{E}_{-1},\L)$. Following this way, one may control the behavior of wave functions of the under-barrier states, which mostly impact the tunneling dynamics. Otherwise, within the standard approach, one needs to modify, in this or those manner, the HO wave functions to fit them to the phenomenological potentials; it results in the unavoidable extension of the number of basic states.

\ssk
To compute the energies and the Floquet wave functions for Hamiltonian \rf{Ht0}, we use states of the exactly-solvable model with the potential  $\tilde{U}(\xi,\bar{E}_{-2},\bar{E}_{-1},\L)$. The parameters of the driving external force are chosen to be close to that of satisfying the CTD condition within the two-level approximation (in the symmetric case) \cite{Grossmann:1992,Llorente:1992,Longhi:2005}:
\be
J_0 \left(\fr{2S r}{w}\right)=0, \qquad r=\langle 1|\xi|2 \rangle .
\la{CTD2lcond}
\ee 
Here $J_0(x)$ is the zero-order Bessel function; $|1\rangle$ and $|2\rangle$ are the lowest states of the undriven system.
Zeros of the Bessel function correspond to the parameters, upon which the tunneling is completely suppressed. 
The resulted Floquet energies for the exactly-solvable Hamiltonian in an external periodic driving force with symmetric and asymmetric two-well potentials are collected in Table 3. 
\begin{table}[ht]
\footnotesize
\centering 
\begin{tabular}{c | c c c c c } 
\hline\hline
Parameters & $N$=0 & $N$=1 & $N$=2 & $N$=3 & $N$=4  \\ [0.5ex] 
\hline
$\L=1$\\ 
$S$=0.65, $w$=0.9 & 0.39038 & 0.358714 & 0.302197 & -0.291837 &-0.189549 \\ [0.5ex] 
\hline\hline
$\L=0.5$\\
$S$=0.65, $w$=0.9 & 0.389543 & 0.358666 & 0.30244 & -0.291951 & -0.188954 \\ [1ex] 
\hline\hline 
Parameters &  $N$=5 & $N$=6 & $N$=7 & $N$=8 & $N$=9 \\ [0.5ex] 
\hline 
$\L=1$\\ 
$S$=0.65, $w$=0.9 & 0.181939 & 0.181259 & -0.119865 & 0.0367728 & 0.0299879  \\ 
[0.5ex]  \hline\hline  $\L=0.5$\\
$S$=0.65, $w$=0.9 & 0.185352 & 0.177982 & -0.120192 & 0.0377385 & 0.0293748  \\ [1ex] 
\hline \hline 
\end{tabular}
\label{table:energyF} 
\caption{The Floquet energies for the symmetric ($\L=1$) and asymmetric ($\L=0.5$) two-well potentials: first ten levels.}
\end{table}
\noindent We restricted the basis with 10 basic wave functions $\vf_k(x)$ entering eq. \rf{Fourier};
extending the number of the involved basic states does not essentially change the results. Note that the level ``numbers'' $N=0,\dots 9$ do not sequentially numerate states from the ground to excited, that is common for the Floquet states (see, for instance, Sec 3.1. of \cite{Grossmann:1991ZP}). However, we can always determine the tunnel doublet following two criteria: the difference between the quasi-energies of states has to be small; the levels have to be (quasi)crossed. Inspecting the quasi-energies of different levels on the plot (see Fig.\ref{fig:image8} as an example), one may notice that the smallest difference is between levels $N=5,6$ of Table 3, and they are (quasi)crossed (see Fig.\ref{fig:image9}). Therefore, just these levels form the tunnel doublet.

\begin{figure}[h!]
\center{\includegraphics[width=0.8\linewidth]{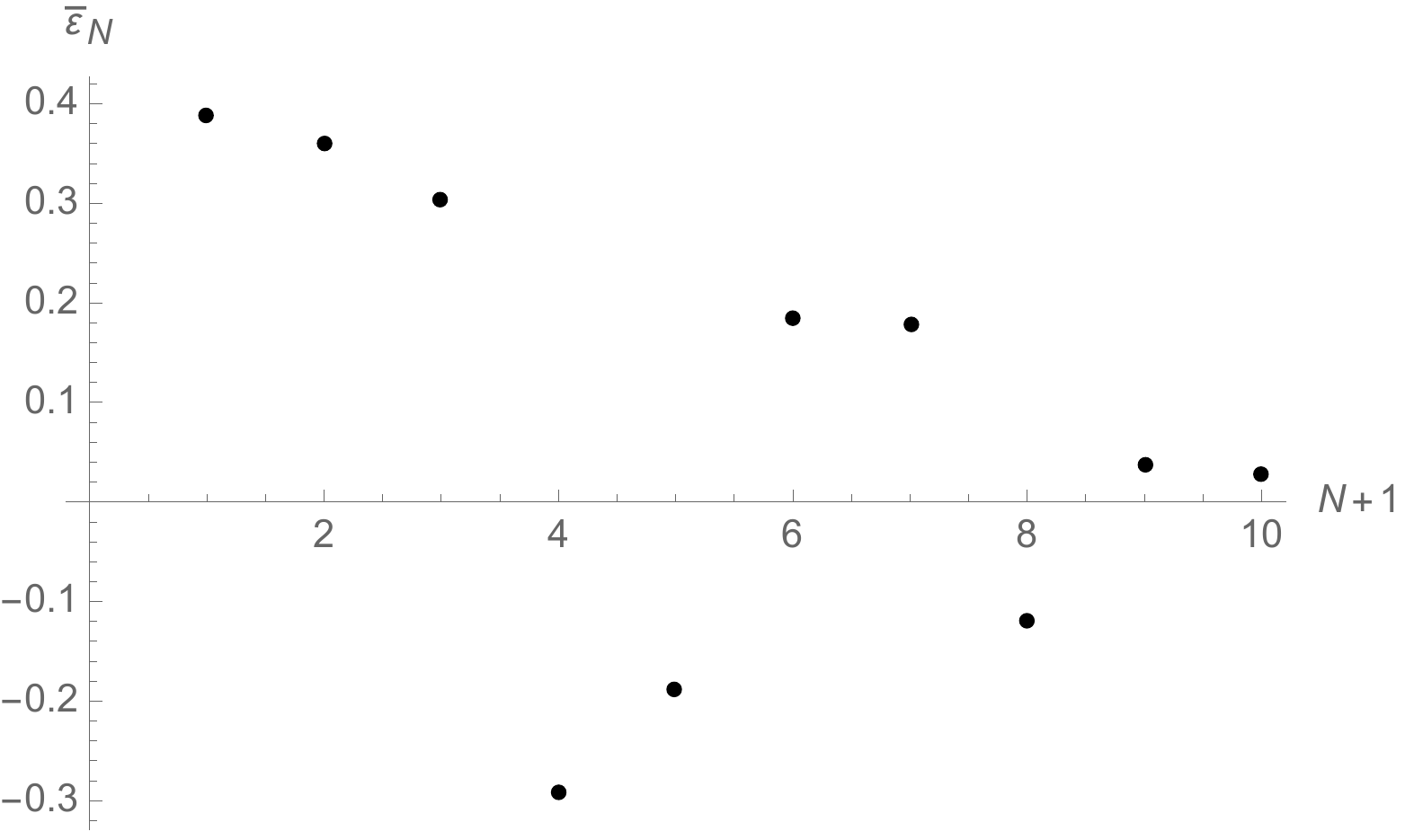}}
\caption{The quasi-energies of Table 3 for the asymmetric two-well potential. The horizontal axis numerates the states; the vertical axis corresponds to the quasi-energies values in $\w_0$ units ($\bar\ve=\ve/\w_0$). The closest quasi-energies correspond to levels $N=5,6$ and $N=8,9$. However, $(\bar{\ve}_5-\bar{\ve}_6)<(\bar{\ve}_8-\bar{\ve}_9)$.}
\label{fig:image8}
\end{figure}

\begin{figure}[h!]
\hspace{-1.cm}
\begin{minipage}[h]{0.499\linewidth}
\center{\includegraphics[width=0.98\linewidth]{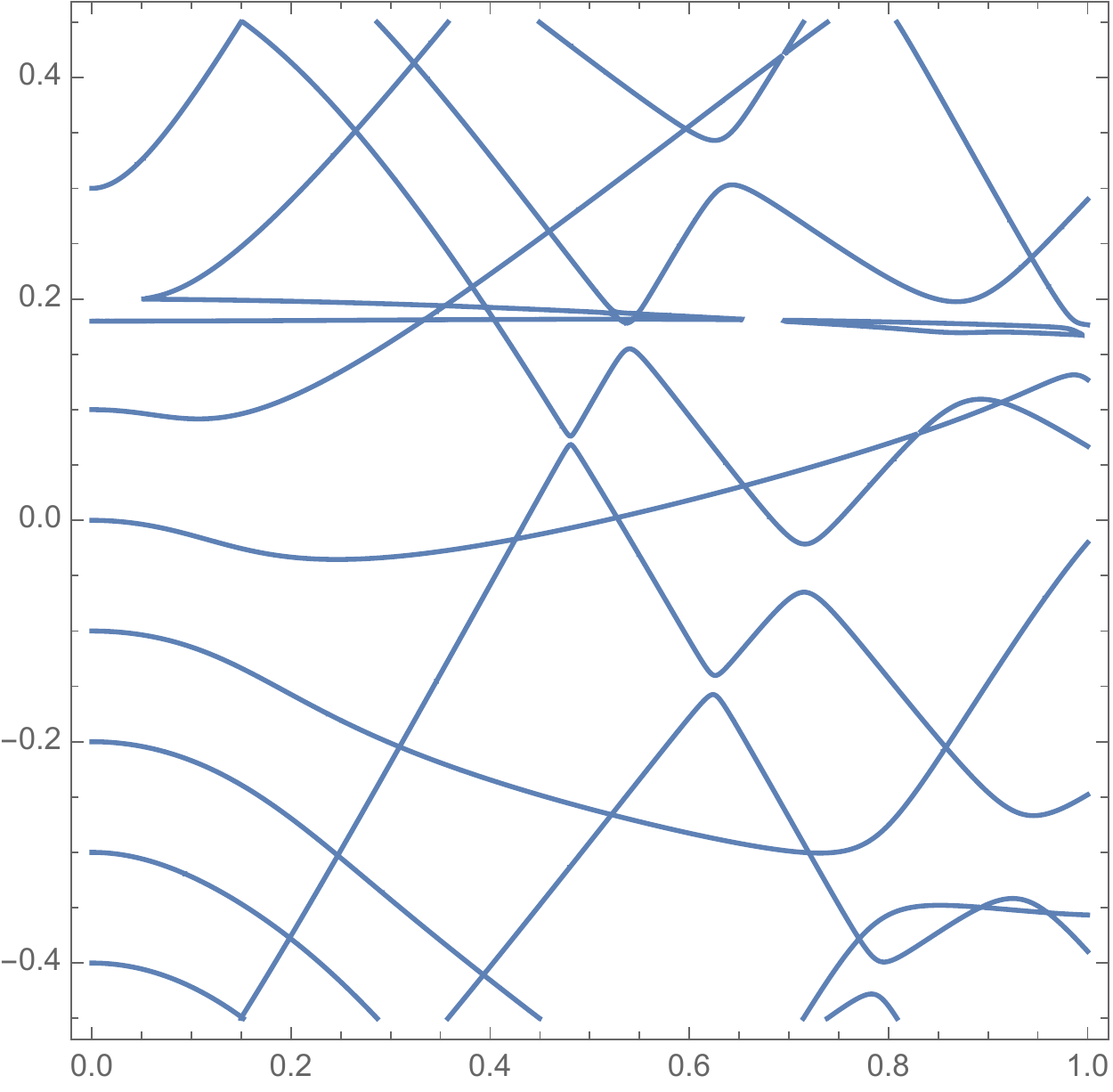} \\ a)}
\end{minipage}
\hspace{1.cm}
\begin{minipage}[h]{0.499\linewidth}
\center{\includegraphics[width=0.98\linewidth]{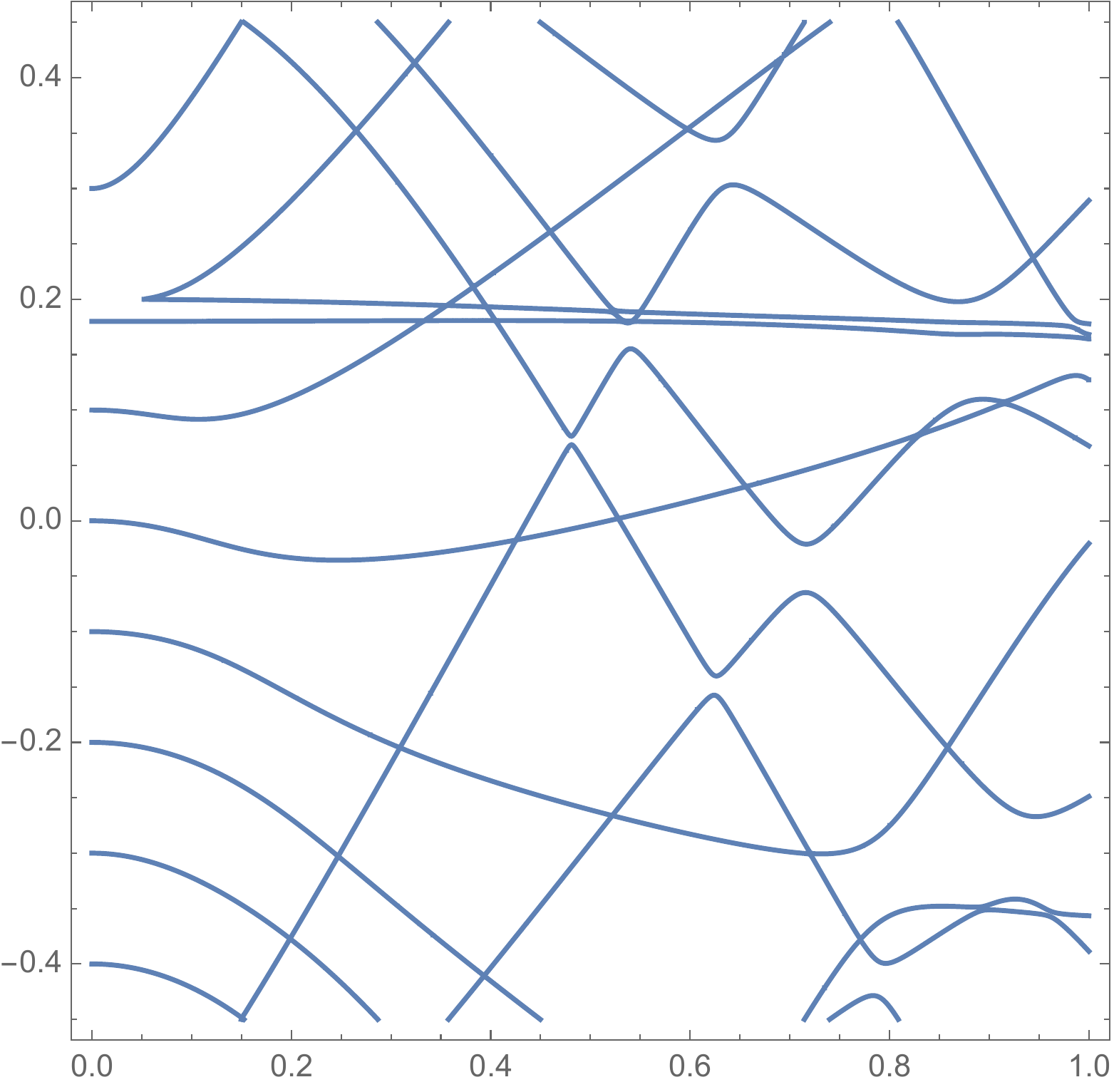} \\ b)}
\end{minipage}
\caption{The levels (quasi)-crossing diagram inside the fundamental domain $\{-w/2,w/2\}$ (vertical axis; $w=0.9$): a) the symmetric case with $\L=1$; b) the asymmetric case with $\L=0.5$. In both cases $S$ varies over zero to one (horizontal axis).
}
\label{fig:image9}
\end{figure}

\ssk
Now one may notice the essential change in the tunnel doublets from their initial value $\triangle=2\cdot 10^{-2}$. In the symmetric case, with the deformation parameter $\L=1$, we get $\triangle^F_{sym}=0.181939-0.181259\approx 6.8 \cdot 10^{-4}$, while in the asymmetric case, with $\L=0.5$, we have $\triangle^F_{asym}=0.185352-0.177982\approx 7.37 \cdot 10^{-3}$. Also, the parameters of the driving field we used are slightly different from that of used in eq. \rf{CTD2lcond}: according to the CTD criterium, in the two-level approximation, one exactly needs $w_{sym}\approx 0.802268$ in the symmetric case and $w_{asym} \approx 0.762012$ in the asymmetric case; $w=0.9$ has been chosen in our computations with taking into account the contribution of more high states of the initial Hamiltonian. However, the proximity of the chosen by us value of $w$ to the values of $w_{sym}$ and $w_{asym}$ justifies the usage of the two-level approximation, at least in the quantitative analysis of the tasks under consideration.

Introducing the so-called stroboscopic variables, $t=nT$, where $T=2\pi/\w$ is the period of the driving force oscillations, essentially simplifies studies of quantum states dynamics.
By use of such variables we can compute the dynamics of a wave packet \rf{Floquetpsi}, expanded over the basis of the Floquet states and energies computed before. As the initial state of eq. \rf{Floquetpsi} we choose the Gaussian wave packet \rf{wavepack}
with different values of the parameters $x_0$ (or $\xi_0$) and $R$. (Recall, we initially set the packet in the right well.) The temporal dynamics of the packet and the parameters of computations can be viewed in Fig.\ref{fig:image10}.

\begin{figure}[h!]
\hspace{-1.3cm}
\begin{minipage}[h]{0.499\linewidth}
\center{\includegraphics[width=0.9\linewidth]{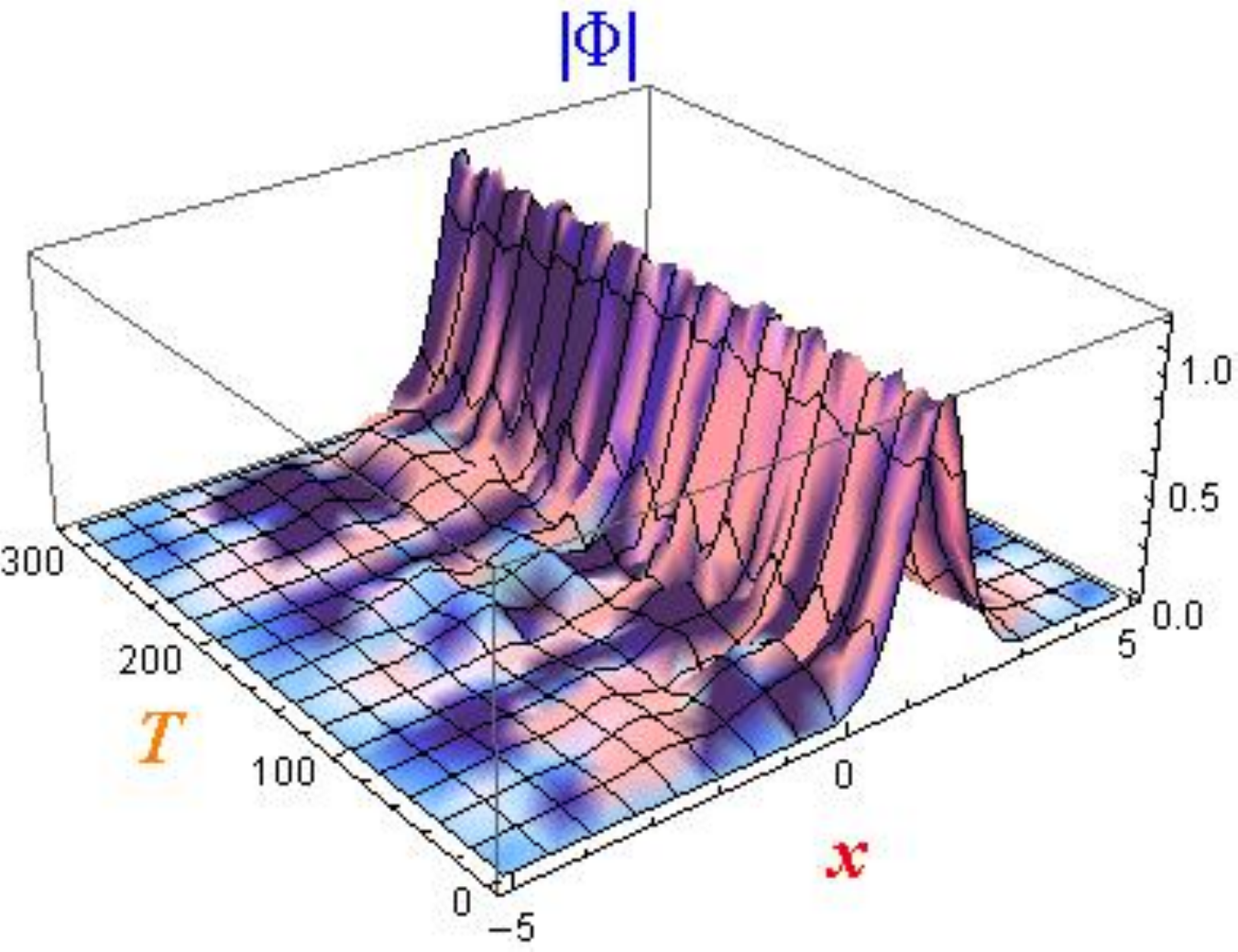} \\ a)}
\end{minipage}
\hspace{1.29cm}
\begin{minipage}[h]{0.499\linewidth}
\center{\includegraphics[width=0.98\linewidth]{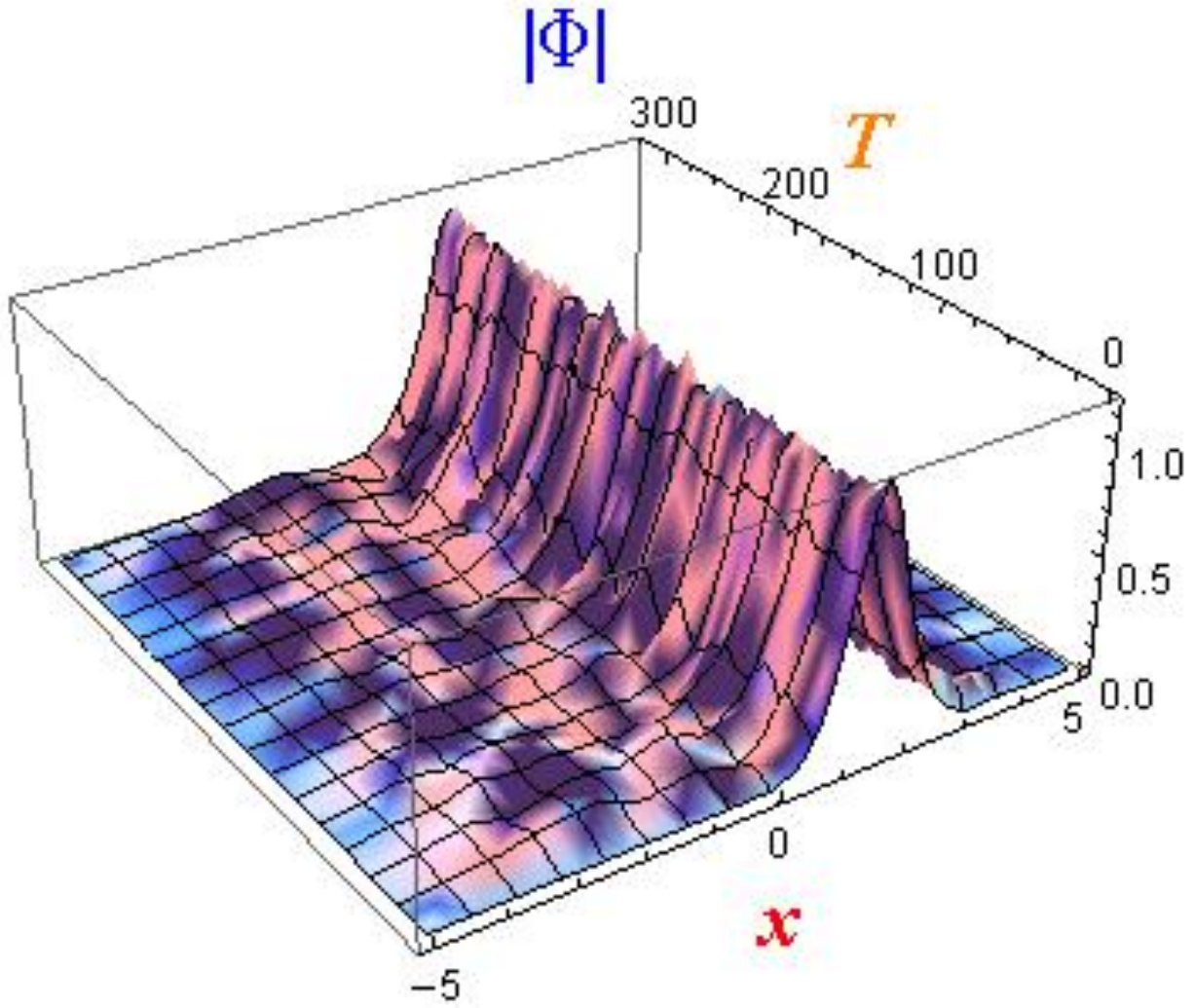} \\ b)}
\end{minipage}
\caption{Evolution of the Gaussian wave packet $\Phi(x,t)$ in the external periodic driving force: a) for the symmetric two-well potential with the initial tunnel doublet $\triangle =0.02$; $R=0.75$, $x_0=1.525$; b) for the asymmetric potential with $\triangle =0.02$ and $\L=0.5$; $R=0.75$, $x_0=1.607$. 
}
\label{fig:image10}
\end{figure} 

\ssk
Looking at Fig.\ref{fig:image10}, one may notice that the wave packet fraction in the left well is negligible in both, symmetric and asymmetric, cases. In other words, wave packets are mainly localized within the initial local minima. There is a difference in localization for asymmetric potentials, since it is provided by a particular localization from the unperturbed case and by the tunneling suppression due to quasi-crossing of the lowest Floquet levels upon turning on the periodic driving. In the symmetric case localization is provided by crossing the Floquet levels of the periodically-driven Hamiltonian \cite{Grossmann:1992}. One may identify a small fracture of the wave packet in the left part of Fig.\ref{fig:image10}a, related to the contribution of higher states.

\section{Summary and Conclusions}

Studies of the tunneling dynamics by use of exactly-solvable models with multi-well potentials is one of examples of the so-called quantum-me\-cha\-ni\-cal engineering. Variations of the reviewed model parameters (value of energies of added states, deformation parameters etc.) allows one to investigate features of the tunneling processes in systems with different number of local minima, and to control the change of their specific parameters (the minima depth, the barriers width, the deformation degree). Recall that the asymptotic behavior of the considered here potentials does not change upon altering the amount of the local minima; models with the phenomenological type potentials do not share such a hallmark. This 
fact might be crucial in studies of the Bose-condensate dynamics of atoms in the Pauli-type traps with several local minima in their central parts. Moreover, we have shown that changing the values of energies of added states still does not go beyond two- or/and three-level approximations, the corresponding parameters of which have been computed within the approach of N=4 supersymmetric quantum mechanics (SQM).

\ssk
Treating the models with or without reflection symmetry within the consistent unified consideration may be regarded as a key advantage of the employed technique. Viz., we have employed N=4 SQM to examine properties of the tunneling processes in models with two- or three-well potentials, induced by a partial localization of levels. The complete localization of levels is unattainable within the followed approach, because it appears upon the considerable enlarging the difference in energies of the tunnel doublet. The latter can not be reached for isospectral Hamiltonians with symmetric or asymmetric potentials. However, turning on a special, ref. \cite{Graffi:1984,Landsman:2013}, breaking the reflection symmetry, perturbation provides the complete localization of the under-barrier states in the considered here examples of multi-well potentials. As we have highlighted, the tunnel doublet gets significantly changed for two-well potentials, when its value (on account of the breaking symmetry term) gets increased in order (in compare to the initial quantity of the tunnel doublet). We have also noted the specifics in localization of levels in models with a three-well potential, where the tunnel doublet is formed by the first and the second excited states. Localization of these states is occurred in outer wells, while the ground state is placed in the central well.

\ssk
The performed studies of the coherent tunneling destruction by a periodically-driven interaction for models with broken reflective symmetry have shown that the criterium for this phenomenon to arise is in agreement with that of a symmetric case. Wave functions and the Floquet spectra are obtained in our approach upon the complete Hamiltonian diagonalization on account of 30 states, in contrast to the symmetric case of ref. \cite{Grossmann:1991}, where to diagonalize the Hamiltonian more than 100 states of the harmonic oscillator have been used. This observation points out one of the repeatedly mentioned throughout the paper advantages of the developed approach, viz., engaging a smart basis of states, that essentially simplifies the computations.

\ssk
Finally, let us conclude with the following point. Partial or complete localization of states is an outcome of playing with free parameters of the model. However, in another language, it may be considered as changing the geometry of an effective internal manifold of quantum states by changing its quantum metric \cite{Kolodrubetz:2013,Ozawa:2018,Ozawa:2019,Bleu:2018}. We believe that applying ideas of the quantum metric approach in exactly-solvable models will give new prospects in studies various properties of quantum systems, the tunneling dynamics as well.

\section*{Acknowledgements}

The authors are thankful to an anonymous reviewer for suggestions in improving the early version of the paper. We also grateful to S.S. Zub for a technical support.


\end{document}